\documentclass[dvips]{acta}
\usepackage{supertabular,lscape,epsfig}
\usepackage{amssymb}
\usepackage{amsmath}
\usepackage{placeins}

\def\degr{\hbox{$^\circ$}}
\def\arcmin{\hbox{$^\prime$}}
\def\arcsec{\hbox{$^{\prime\prime}$}}

\def\farcm{\hbox{$.\mkern-4mu^\prime$}}

\SetPages{0}{0}

\SetVol{0}{0}

\begin{document}

\begin{Titlepage}
\Title{1.4-GHz observations of extended giant radio galaxies\footnote{Based on the observations with the 100-m telescope at Effelsberg
operated by the Max-Planck-Institut f\"ur Radioastronomie (MPIfR) on behalf of the Max-Planck-Gesellschaft.}}
\Author{We\.zgowiec, M.}{Obserwatorium Astronomiczne Uniwersytetu Jagiello\'nskiego, ul. Orla 171, PL-30244 Krak\'ow, Poland\\
e-mail:markmet@oa.uj.edu.pl}

\Author{Jamrozy, M.}{Obserwatorium Astronomiczne Uniwersytetu Jagiello\'nskiego, ul. Orla 171, PL-30244 Krak\'ow, Poland\\
e-mail:jamrozy@oa.uj.edu.pl}

\Author{Mack, K.-H.}{INAF -- Istituto di Radioastronomia, Via P. Gobetti 101, I-40129 Bologna, Italy\\
e-mail:mack@ira.inaf.it}

\Received{}
\end{Titlepage}

\Abstract{This paper presents 1.4-GHz radio continuum observations
of 15 very extended radio galaxies. These sources are so large that
most interferometers lose partly their structure and total flux density. Therefore,
single-dish detections are required to fill in the central (u,v) gap of interferometric data
and obtain reliable spectral index patterns across the structures, and thus also an integrated
radio continuum spectrum. We have obtained such 1.4-GHz maps with the 100-m
Effelsberg telescope and combined them with the corresponding maps available from the NVSS.
The aggregated data allow us to produce high-quality images, which can be used to
obtain physical parameters of the mapped sources. The combined images reveal in many cases extended
low surface-brightness cocoons.}{galaxies: active -- galaxies: evolution -- galaxies: jets}

\section{Introduction}
The classic powerful radio galaxies (RGs) are characterized by extended radio
lobes with leading compact and bright hot spots, and often a compact central
radio core. The lobes are powered by two relativistic jets emerging from a
supermassive black hole at the centre of a galaxy (e.g.~Scheuer~1974). There
is a huge diversity in linear extent of RGs: from less than $10^{2}$~pc --
Gigahertz-Peaked Spectrum (GPS), $10^{2} - 10^{4}$~pc -- Compact Steep Spectrum (CSS),
up to $10^{4} - 10^{6}$~pc -- normal size sources. Due to instrumental limitations it
had taken some time before it was realized that there were some extremely extended extragalactic
objects nearby. The first of them was 3C\,236, discovered by Willis et al.~(1974), which 
with its $\sim$ 4 Mpc size, for more than four decades was the record holder.
Ishwara-Chandra \& Saikia~(1999) presented a sample of all 50 known giant radio galaxies (GRGs)
that were published in the literature up to the beginning of the new millennium.
The use of the new ``all-sky'' radio surveys performed by the Westerbork Synthesis Radio Telescope (WSRT),
the Very Large Array (VLA), and the Australia Telescope Compact Array (ATCA) interferometers
played an important role in finding structures of Mpc sizes. Indeed, owning to the Westerbork
Northern Sky Survey (WENSS, Rengelink et al.~1997) at 0.3\,GHz, the Sydney University Molonglo
Sky Survey (SUMSS, Bock et al.~1999) at 0.8\,GHz, the NRAO VLA Sky Survey (NVSS, Condon et al.~1998)
and the Faint Images of the Radio Sky at Twenty-Centimeters (FIRST, Becker et al.~1995)
at 1.4\,GHz, several dozens of new GRGs were found over the entire sky. The new samples by
Schoenmakers et al.~(2000), Saripalli et al.~(2005), Lara et al.~(2001), and Machalski et al.~(2001, 2006, and 2007), 
as well as Ku\'zmicz and Jamrozy~(2012) raised the number of known GRGs to about 300 objects. GRGs, which are the
largest single objects in the Universe, are believed to have evolved from the smaller sources,
however the details are not yet fully understood. 
There are a number of factors considered to underlie such a large size of these radio sources.
In the paper by Subrahmanyan et al.~(1996), which presents research on several bright giant radio galaxies,
it is postulated that the giants could become so large after several jet activity periods of their central AGN.
Therefore, some of the very extended radio sources may be surrounded by remnants of their former jet activity.
GRGs are extremely useful for studying a number of astrophysical problems. These range from understanding the evolution
of radio sources, constraining orientation-dependent unified schemes, to probing the intergalactic
(IGM) and intracluster (ICM) media at different redshifts.

There are two basic types of extragalactic radio sources: FRI and FRII (Fanaroff and Riley~1974),
which differ mainly morphologically, as well as in their radio luminosities
(the threshold value is $L_{178\,MHz}\sim 2 \times 10^{25}\,W Hz^{-1} sr^{-1}$, for H$_0$ = 50\,km s$^{-1}$ Mpc$^{-1}$).
The luminosity boundary between them, however, is not too sharp, and there is some overlap in the luminosities
of sources classified as FRI or FRII on the basis of their morphology.
FRI sources are less bright objects with diffuse radio lobes, whose brightest parts are situated close to the nucleus.
Their radio brightness decreases with distance from the central source. In the radio lobes of FRII radio galaxies,
the brightest parts are located at their ends, forming so-called hot spots, i.e. places where a jet of relativistic matter
collides with the intergalactic medium, creating a supersonic shock wave.
The plasma flowing back from the terminal shocks forms a bridge between the lobes.
This diffuse emission in the central parts of FRII sources comes from the oldest population of charged particles.
The large-scale jets in FRII type radio galaxies are better collimated, although they are more asymmetric than those in FRIs.
These asymmetries are due to the Doppler beaming of jets moving towards the observer.
It is possible that at least some of the FRI radio galaxies began as FRIIs in dense environments of rich clusters
and relatively quickly evolved to the FRIs (Hill and Lilly~1991). However, there are also hybrid morphology radio galaxies (HyMoRS),
which show different Fanaroff-Riley radio structure on either side of the active nucleus (Gopal-Krishna and Wiita~2000).
The existence of HyMoRS seems to contradict the both types having an evolutionary connection.

The large angular sizes of many GRGs (at least of those located at low redshifts) provide a great
opportunity to study their physical conditions in several different locations (independent
points) within their lobes. On the other hand, the large angular size of GRGs is also a
disadvantage, as currently available instruments do not possess a sufficient dynamic range to
allow detections of faint diffuse structures and strongly emitting features in the same field.
These limitations mean that there is a residual flux (mostly negative) near
strong sources, many times greater than the nominal rms background
of $\sim$0.5\,mJy/beam.
Moreover, interferometric observations of extended RGs are characterized by poor sensitivity
to extended structures, due to the so-called ``missing zero spacing'' effect: the larger the
extent of an observed source, the bigger both structure and flux-density losses. For instance,
the VLA operating in its most compact configuration is insensitive to structures 
larger than about $16\arcmin$ at the frequency of 1.4\,GHz. Jamrozy et al.~(2004) demonstrated how severe that
problem can be in the case of a fossil radio galaxy B2\,0924+30, where about 30\% of the flux density
(and structure) is lost in the interferometric map obtained from the NVSS.

\noindent
The best way to overcome the problems mentioned above is to merge the high-resolution interferometric
data with single-dish observations. This allows to
combine the short scales of interferometric observations that provide angular resolution with the large scales
of single-dish data, assuring high sensitivity to the entire extended radio emission. The single-dish data
recover the flux lost due to both ``missing zero spacing'' and low dynamic range of the interferometer.
In this paper we show our results from combining single-dish maps with interferometric data. 
The observations at 1.4 GHz for a sample of the most extended GRGs
located in the northern hemisphere were performed by the 100-m Effelsberg telescope while the interferometric data
were taken from the NVSS survey.

As all sources presented in this paper are well-known objects, comprehensively studied in different domains, we focus
on recovering a possible diffuse large-scale emission with the use of sensitive single-dish observations and discuss the results
in terms of existence of extended halos/cocoons around the giant radio galaxies under study. This method of data combination has never been
used for this class of radio sources and we are going to show that it is very important in studying of morphologies and evolution
of FRI and FRII radio galaxies.

In Sect.~2, we describe the observed sample and give the details of observations and data reduction, including
the description of data combination process. In Sect.~3, we present our results and discuss them in Sect.~4.

\begin{landscape}
\begin{table}
\caption{\label{objects}Physical details of the studied GRGs and parameters of the 1.4\,GHz Effelsberg observations}
\begin{center}
\begin{tabular}{llccccrcccrr}
\hline\hline
Source          &Other          &\multicolumn{2}{c}{Field centre}       & Ang.                          & $z$                   &  FR           & Lin.  &Map            &Time/  &No.    &r.m.s \\
                &name           & R.A.$_{J2000}$& DEC.$_{J2000}$        & size                          &                       & type          & size  &size           &cover. &of     & [mJy/ \\
                &               &[$\rm^{h~m~s}$]&[$\degr~\arcmin~\arcsec$]&[$\arcmin$]                  &                       &               & [Mpc] &$[\arcmin]$    &[min]  &cov.   &  b.a.]\\
\hline
J0057+3021      & NGC\,315      & 00 57 49   & $+$30 21 09              & 58.0$^{\rm a}$         & 0.0167$^{\rm a}$& I/II         & 1.17  & 96            & 25    &  6    & 11.6 \\
J0107+3224      & 3C\,31        & 01 07 25   & $+$32 24 45              & 40.0                   & 0.0170$^{\rm b}$&   I          & 0.82  & 78            & 18    & 10    & 11.0 \\
J0318+6829      &               & 03 18 19   & $+$68 29 31              & 14.9$^{\rm a}$         & 0.0902$^{\rm a}$&  II          & 1.49  & 48            & 10    &  6    & 10.1 \\
J0448+4502      & 3C\,129       & 04 48 58   & $+$45 02 01              & 30.0$^{\rm a}$         & 0.0210$^{\rm a}$&   I          & 0.75  & 66            & 15    &  6    & 12.7 \\
J0702+4859      &               & 07 02 06   & $+$48 59 17              & 19.1                   & 0.0650$^{\rm c}$& I/II         & 1.41  & 54            & 11    &  8    &  6.7 \\
J0748+5548      & DA\,240       & 07 48 37   & $+$55 48 58              & 34.0$^{\rm a}$         & 0.0356$^{\rm a}$& I/II         & 1.43  & 72            & 16    &  6    & 13.5 \\
J0949+7314      & 4C\,73.08     & 09 49 46   & $+$73 14 23              & 14.7$^{\rm a}$         & 0.0581$^{\rm a}$& I/II         & 0.98  & 54            & 11    &  6    &  5.9 \\
J1006+3454      & 3C\,236       & 10 06 02   & $+$34 54 10              & 39.0$^{\rm a}$         & 0.0988$^{\rm a}$&  II          & 4.22  & 78            & 18    &  8    & 11.5 \\
J1032+5644      & HB\,13        & 10 32 59   & $+$56 44 53              & 35.0$^{\rm a}$         & 0.0450$^{\rm d}$&  I           & 1.83  & 54            & 11    &  8    &  7.1 \\
J1312+4450      &               & 13 12 17   & $+$44 50 21              & 22.6                   & 0.0358$^{\rm c}$& I/II         & 0.95  & 60            & 12    &  8    &  9.4 \\
J1428+2918      &               & 14 28 19   & $+$29 18 43              & 14.7                   & 0.0870$^{\rm c}$&  II          & 1.42  & 54            & 11    &  6    &  6.4 \\
J1552+2005      & 3C\,326       & 15 52 09   & $+$20 05 24              & 19.6$^{\rm a}$         & 0.0895$^{\rm a}$&  II          & 1.94  & 60            & 12    &  7    & 12.8 \\
J1628+5146      & MRK\,1498     & 16 28 05   & $+$51 46 31              & 19.0                   & 0.0560$^{\rm a}$&  II          & 1.22  & 54            & 11    &  6    & 12.9 \\
J1632+8232      & NGC\,6251     & 16 32 32   & $+$82 32 17              & 52.0$^{\rm a}$         & 0.0230$^{\rm a}$& I/II         & 1.43  & 90            & 23    &  6    & 10.5 \\
J2145+8154      &               & 21 45 31   & $+$81 54 54              & 18.3                   & 0.1457$^{\rm c}$&  II          & 2.77  & 54            & 11    &  8    &  5.4 \\
\hline
\end{tabular}
\end{center}
For the linear size we use a Hubble constant of $H_0=71\,km/s/Mpc$ (with $\Omega_{M} = 0.27$ and $\Omega_{\Lambda} = 0.73$); Spergel et al.~(2003).\\
$^{\rm a}$Ishwara-Chandra and Saikia~(1999).\\
$^{\rm b}${\it redshift}: Smith et al.~(2000); {\it angular size}: Andernach et al.~(1992).\\
$^{\rm c}$Schoenmakers et al.~(2001).\\
$^{\rm d}${\it redshift}: Falco et al.~(1999); {\it angular size}: Masson~(1979).\\
\end{table}
\end{landscape}

\section{Sample, observations, and data reduction}
\label{sample}
We selected a complete sample of 15 GRGs (see Table~1 for details) that have
angular sizes greater than $14\farcm5$ and located at declinations $\delta>0\degr$.
We mapped the targets at 1.4~GHz in total power with the 100-m single-dish Effelsberg telescope
in order to fill in the missing short spacings of the VLA data from the NVSS survey.

The single-dish measurements were carried out with the Effelsberg
telescope during the second half of 2004, starting at JD\,2453234.159722,
JD\,2453235.161111, and JD\,2453259.752083. The project number was 67-04. We used
a single horn, 2-channel receiver installed in the primary focus of the telescope.
The final images were combinations of individual maps scanned alternatively
in right ascension and declination. A total of about 7 coverages for each target
were obtained. The maps sizes were chosen to cover the expected radio extent of the target,
as well as some emission-free areas used for determination of zero levels and noise.
The scanning velocity of the antenna was 3\degr/min and the scan interval 3\arcmin
(to avoid aliasing).

The VLA interferometric data were acquired from the NVSS~(Condon et al.~1998), which is
a set of $\sim$2300 1.4 GHz continuum maps covering 4$^{\rm o} \times 4^{\rm o}$ each.
They cover the entire sky north of declination
-40$^{\rm o}$. The angular resolution is $45\arcsec \times 45\arcsec$ and the r.m.s.
is about 0.45 mJy/beam ($\approx$0.14 K) in total intensity.
The r.m.s. is somewhat larger near strong sources, whose sidelobe responses
are not completely removed by cleaning. The local dynamic range of the total intensity
images is about 1000:1. The r.m.s uncertainties in right ascension and declination
vary from $\leq1\arcsec$ for sources stronger than 15 mJy to about $7\arcsec$ at the survey limit.
The completeness for point sources of the survey is about 2.5 mJy. It contains over 1.8
million of individual objects. The poor (u, v) coverage of the NVSS snapshot observation makes it
difficult to obtain a good photometic accuracy on source components extended by more than a few
synthesized beamwidths. Moreover, the noise error of the flux density of a faint source is
multiplied by the square root of the number of independent beam areas covered by the source.
Thus the NVSS images are insensitive to smooth radio structures much larger than several
arcminutes in both coordinates. This only shows that adding sensitive single-dish data
can significantly increase the quality of the final maps, as well as provide more precise measurements
of physical parameters of observed sources.

\subsection{Data reduction}
\label{datared}

In this work we used already reduced and calibrated NVSS fits maps that were only geometrically
modified (i.e. re-gridded and re-sized) to match our Effelsberg radio maps for further combination processes.
Therefore, the data reduction performed for the purpose of this work involved the Effelsberg single-dish data only.
Below, we briefly present the main steps of this data reduction and discuss in detail the procedure to combine the VLA and the Effelsberg data used in this paper.

\subsubsection{Effelsberg single-dish data}
\label{eff}

The data reduction  was performed using the NOD2 data reduction
package~(Haslam~1974). All coverages were combined using the spatial frequency
weighting method~(Emerson and Gr\"ave~1988), yielding
the final maps of total power. In order to
remove spatial frequencies corresponding to noisy structures
smaller than the beam size, a digital filtering process was applied
to the final maps. Next, the maps were exported to the fits format and calibrated.
The flux density calibration of the obtained maps was performed using the standard calibration
sources 3C\,286 and 3C\,138, applying the scale of Baars et al.~(1977).
The resulting size of the beam was 9\farcm6. At this step the Effelsberg maps were ready to
be merged with the NVSS interferometric data. The technical details of the maps are given in Table~1.

\subsubsection{Data combination}
\label{imerg}

Combination of the single-dish and NVSS maps has been performed using
{\it IMERG} included in the AIPS package\footnote{http://www.aips.nrao.edu}. This task allows to merge
two input images. As this task uses a Fourier transform to merge images, we needed to make sure that the maps
are of the size of 2$^n$ pixels. For better performance of the computing process, we fixed the size of all maps to
1024 $\times$ 1024 pixels, using tasks {\it LGEOM and HGEOM}. In order to avoid errors of the Fourier transform resulting from blanked pixels
(produced during the process of matching maps), we used a task {\it REMAG}, which puts the zero value for each blanked pixel.
The maps prepared with the recipe described above were then provided to the task {\it IMERG}.
First, a Fourier transform of both maps was carried out, normalizing the low-resolution map (Effelsberg)
to amplitudes within an (u, v) annulus of the high-resolution map (VLA) and then producing an
output transform plane, consisting of the inner plane from the low-resolution map.
The output image was the back transform of this merged (u, v) plane.
The shortest baseline separation for the VLA D-configuration was less
than half of the Effelsberg dish diameter, which helped to determine the
relative scaling of the two data sets (Stanimirovic~2002).
For the minimum baseline a value of 0.1 k$\lambda$ was adopted and
the maximum baseline was a free parameter, though not exceeding 0.5 k$\lambda$.
These baseline numbers defined the annular region (parameter {\it UVRANGE})
of an assumed overlap between the low- and high-resolution image transforms.
The normalization of the low-resolution image is based on comparing the mean amplitudes
in this annulus. For the normalizing factor (parameter {\it FACTOR}) we used a value of 0.006, which is the
ratio of the beam sizes at 1.4~GHz for the VLA D-array and the Effelsberg telescopes.
However, prior to fixing of this value, the parameter was left free and the maximum baseline was chosen in such a way
that the normalizing factor calculated by the {\it IMERG} task was equal or very close to the theoretical value of 0.006.
With the chosen maximum scale fixed, an appropiate size of the annulus of overlap provided the best conditions for image
merging. This way of setting parameters helped to account for slight changes in the (u,v) coverages of the NVSS pointings.

The merged maps were checked for the flux
compliance. To do that, we compared the total flux of the source in the Effelsberg map with the flux of the same area in the VLA merged map. This was
neccessary in order to take into account all point sources unresolved in the Effelsberg map. In all the cases the difference did not exceed 5\%, still within the flux error for all objects (see Table~3). We performed also the same check for the point sources (in the VLA and merged maps).
Here, the compliance is even better, as the flux differences did not exceed 3\%. This close correspondence of the fluxes also suggested that
possible negative-flux artifacts from the NVSS map did not influence significantly the final merged maps, especially that the calculated flux errors were typically
tens or few hundreds of mJy. The final maps lacked significant negative artifacts that were succesfully replaced with the single-dish data. The only exception was
the map of GRG\,J0448+4502, which will be discussed in Sect.~3.4.

A similar technique, called ``feathering'', was used before by Law et al.~(2008) for merging the VLA and GBT
observations at 1.4 and 5 GHz of a field near the Galactic Centre. The work presented here, however, is the first attempt to combine interferometric and
single-dish data for GRGs. As it will be shown later in this paper, for a thorough analysis of the radio emission from such sources,
the use of the above-mentioned technique is crucial. We also note here that the combination of the VLA and the Effelsberg (or the GBT) data provides
the best results, as the shortest baselines in the most compact D-configuration and the resulting ``missing zero spacing''
area perfectly match the size of the single-dish of 100\,meters, allowing also
for an important anullus of overlap in the (u,v) space.

\subsubsection{Sidelobe artifacts}
\label{sidelobes}

Observations of very strong radio sources with a large beam, like in the case of our Effelsberg data, can lead to sidelobe patterns visible as artifacts in the final
maps. Since no ``beam patterns'' at 1.4\,GHz obtained with the same receiver (replaced recently with a more modern one) are available, we cannot ``clean''
the data to remove such artifacts. While we do not see any distinct patterns in the final maps, we estimate the possible influence of the artifacts, as this could lead
to a change of the flux distribution throughout the observed field.
For the beam size of our observations of $9\farcm6$, the first sidelobe would appear $12\arcmin$ from the peak of a point source, having 1.74\% of its intensity. The second one - as much as $20\arcmin$ away with only 0.4\% intensity. Further sidelobes contribute to a neglectible level. Assuming, for simplicity, that the entire measured flux
comes from the central source of the radio galaxy, we can thus conclude that the sidelobes can change the observed flux by about 2\%. As we mentioned above, for
all sources, the accuracy of our integrated flux measurements was 5 to 6\% (also due to amplitude calibration uncertainties). 
So any possible distortions from the sidelobe effects did not influence the
flux measurements to any significant level. Similarly, the absence of visible artifacts in the final merged maps shows that in the case of our study, the sidelobe
contribution can be neglected. 
We would like to note here, however, that the influence of the sidelobes can be considerable at frequencies of the order of 10\,GHz or higher (Klein and Mack~1995).

\section{Results}
\label{results}

Here we present radio maps for all the objects from our sample. For each object, a two-panel figure is presented.
The upper (or left) panel shows the original total-power (TP) maps obtained with the Effelsberg (greyscale) and the VLA (contours) telescopes.
The lower (or right) panel shows the merged TP map (both greyscale and contours), which reveals low surface-brightness extended
structures of a radio source without loosing the high resolution of details inside the lobes.
All the contour maps are plotted with contours of 3, 5, 8, 16, 30, 40, 50, 60, 80, 100, 200 times the noise level given in the
caption for each map individually (and rounded to 0.1~mJy/b.a.). 
The circle in the bottom left corner of each map marks the beam size of the NVSS data. For legibility reasons,
the beam size of the Effelsberg data ($9\farcm6$) is not shown. For each figure, a short description in the relevant section is provided.

The flux densities at 1.4\,GHz presented in Table~3 were obtained by integrating the signal in
polygonal areas encompassing all the visible radio emission in the merged map with exclusion of point sources when possible.
We did not perform source subtraction for point sources residing within the extended emission from the objects, as in all cases the additional
flux was lower than the flux error. Exclusion of point sources would also result in underestimating the actual flux of the sources, which we wanted
to avoid. Below, we comment each of galaxies.

\subsection{\it GRG J0057+3021}
\label{ngc315}

The merged map of NGC\,315 (Fig.~1), similarly as in the case of NGC\,6251, shows the extent and morphology of the source to a
similar degree as the map at 325\,MHz from~Mack et al.~(1997). Although the continuity of the western jet, as well
as its bending are traced more clearly than in the NVSS map, the ``backflow'' part seems to be disrupted, as if the lobe emanated 
directly from the central source. The eastern lobe and its extension to the north has been reproduced to a significant level in the merged
map. Due to both size and morphology of the source, the NVSS map lacks as much as 52\% of the total flux restored from the single-dish map.

\begin{figure}
\begin{center}
\resizebox{\hsize}{!}{\includegraphics[angle=-90]{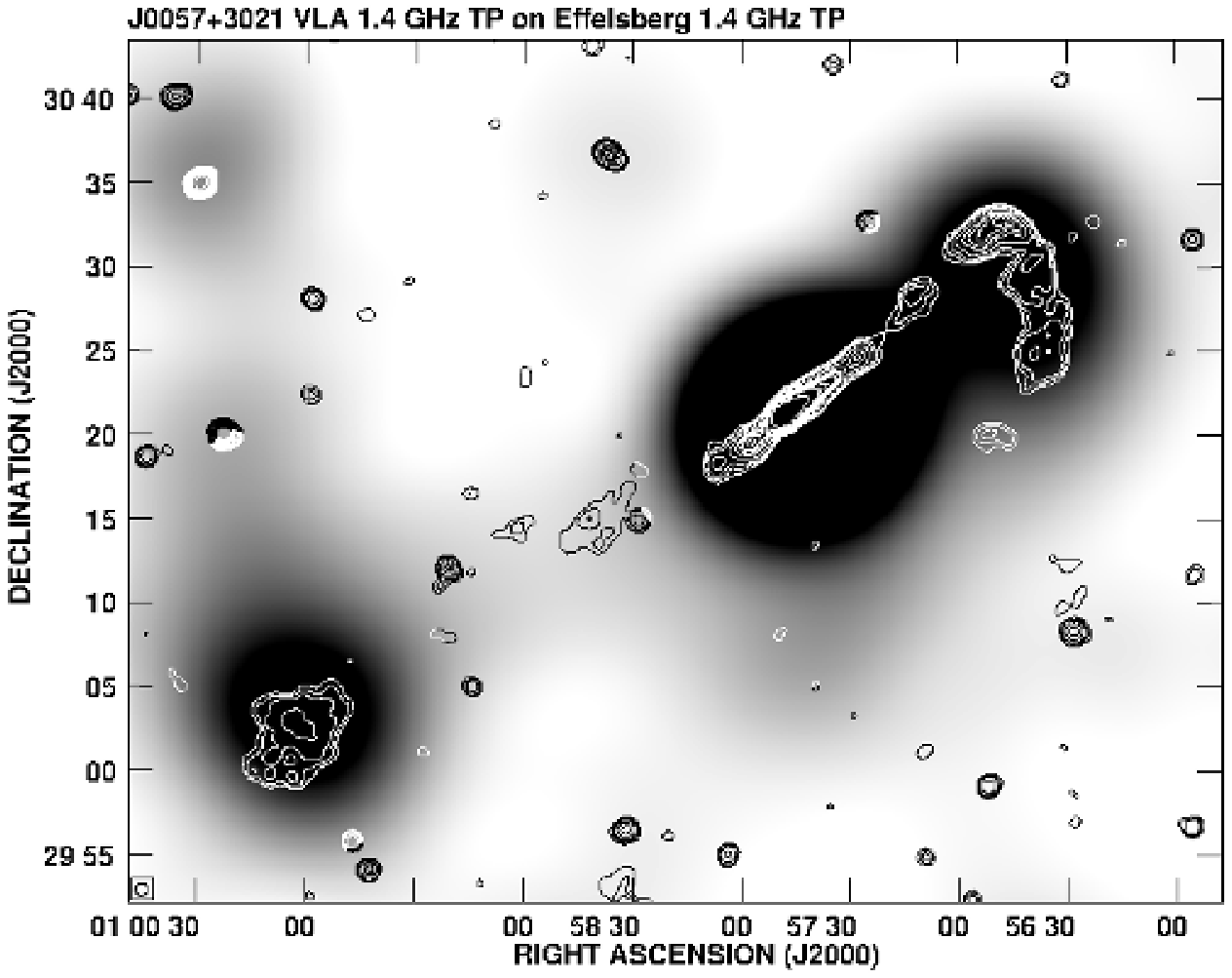}}
\resizebox{\hsize}{!}{\includegraphics[angle=-90]{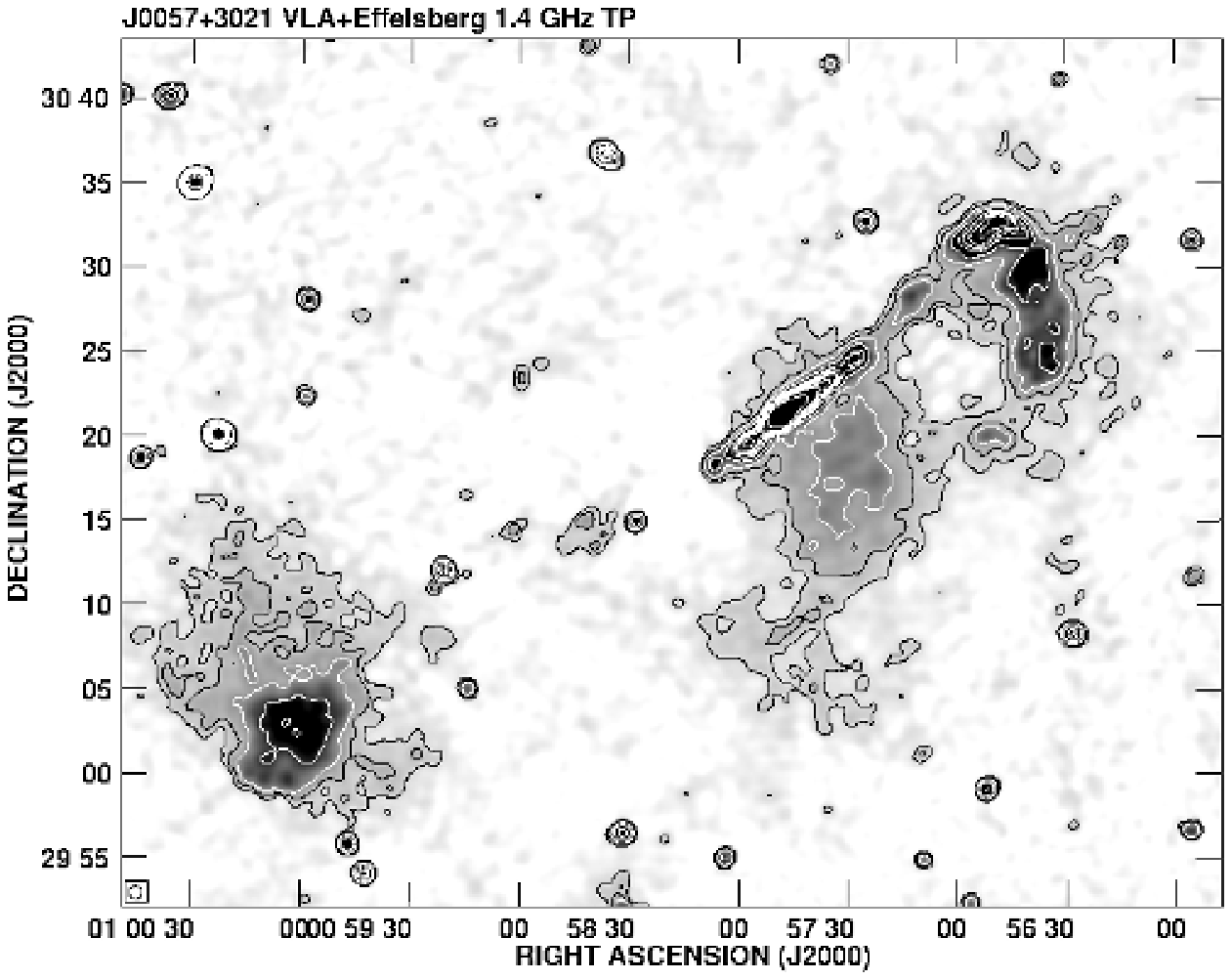}}
\FigCap{GRG J0057+3021. Noise level in both maps is at 0.5~mJy/b.a..}
\end{center}
\label{J0057}
\end{figure}

\subsection{\it GRG J0107+3224}
\label{3c31}

This radio galaxy was extensively studied in the past in many domains while mostly at radio wavelengths.
There are more than 480 publication dealing with this source in the ADS.
3C\,31 was observed in great detail by~Laing et al.~(2008a,2008b), who showed that this radio galaxy produces
complex outflows in the form of multiple jets and tails. Fig.~2 shows that the entire complex is embedded in a diffuse faint radio cocoon.
The appearance of the diffuse structure around the bright north-south oriented radio structure can be a bit surprising,
as it is not visible in any other map known to date. This weak large-scale structure
extends predominantly to the southwest direction of 3C\,31. Artyukh et al.~(1994) show a low-resolution
120\,MHz image of 3C\,31. There is a distinct extension to the west in this map with two unrelated radio sources to be seen within it. Their extrapolated 102\,MHz flux density, however,
does not account for the observed total flux density excess. Furthermore, the analysis of the ROSAT data by~Trussoni et al.~(1997) 
and~Komossa and B{\"o}hringer~(1999) showed an arcmin-scale X-ray emission extending mostly to the southwest from the 3C\,31 host galaxy.
Also~Hardcastle et al.~(2002) noticed an increased X-ray background around 3C\,31 in the Chandra observations.
The authors argued that the soft X-ray emission is mostly associated with the gas within the Arp\,331 group of galaxies.
Taking into account that for the clusters of galaxies which possess radio halos the radio power
of halos correlate with the cluster's X-ray luminosity~(e.g.~Feretti et al.~2012),
it could be possible that also in this case the diffuse X-ray emission from the IGM is related to the radio halo visible in our image. Thus the observed cocoon can be simply a radio halo around the entire cluster, not directly linked to 3C\,31. Such an interpretation would be supported
by the fact that the total flux from our merged map stays above the modelled spectrum (Fig.~16), suggesting some additional flux not associated
with the source.

\begin{figure}
\begin{center}
\resizebox{0.35\vsize}{!}{\includegraphics{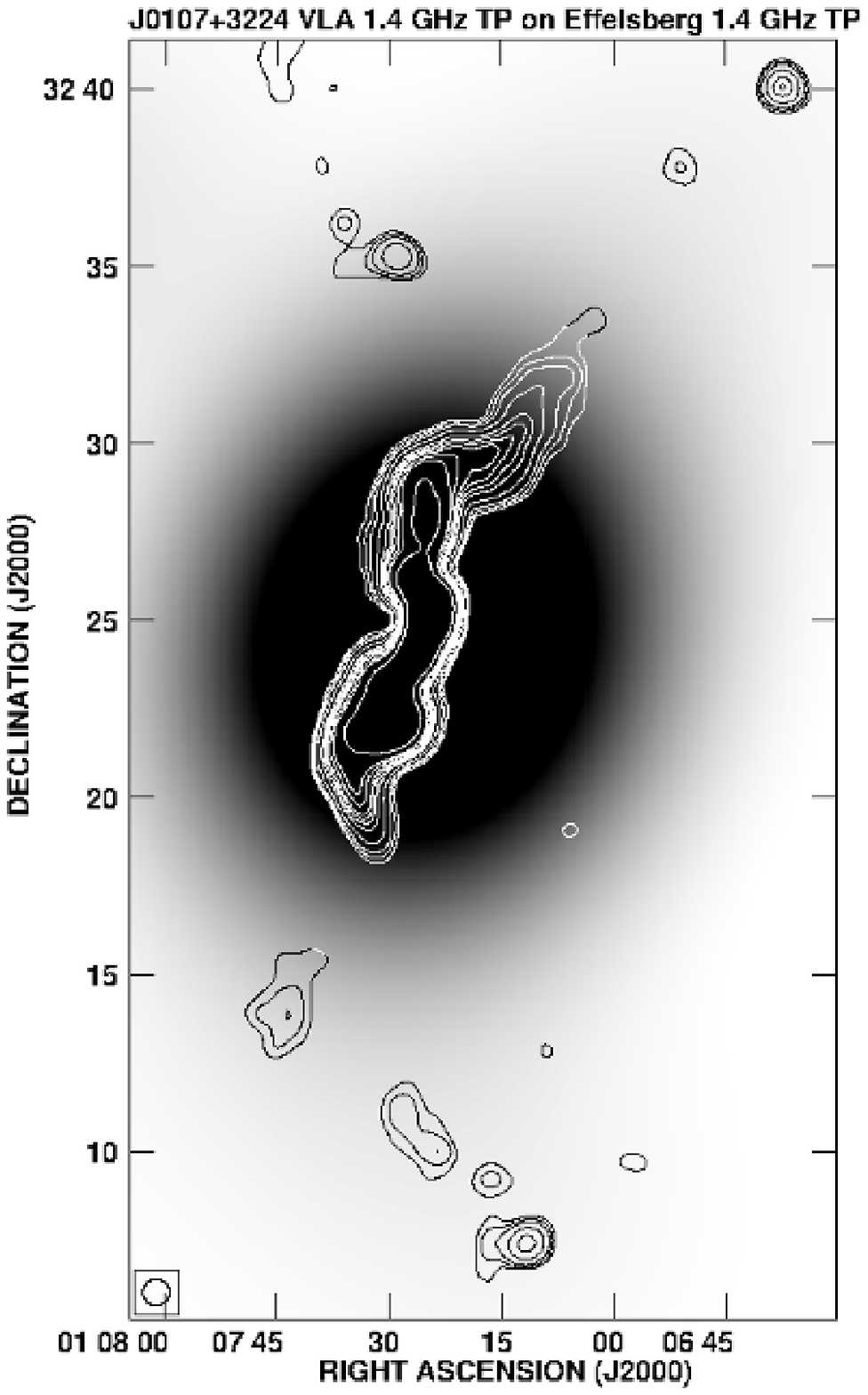}}
\resizebox{0.35\vsize}{!}{\includegraphics{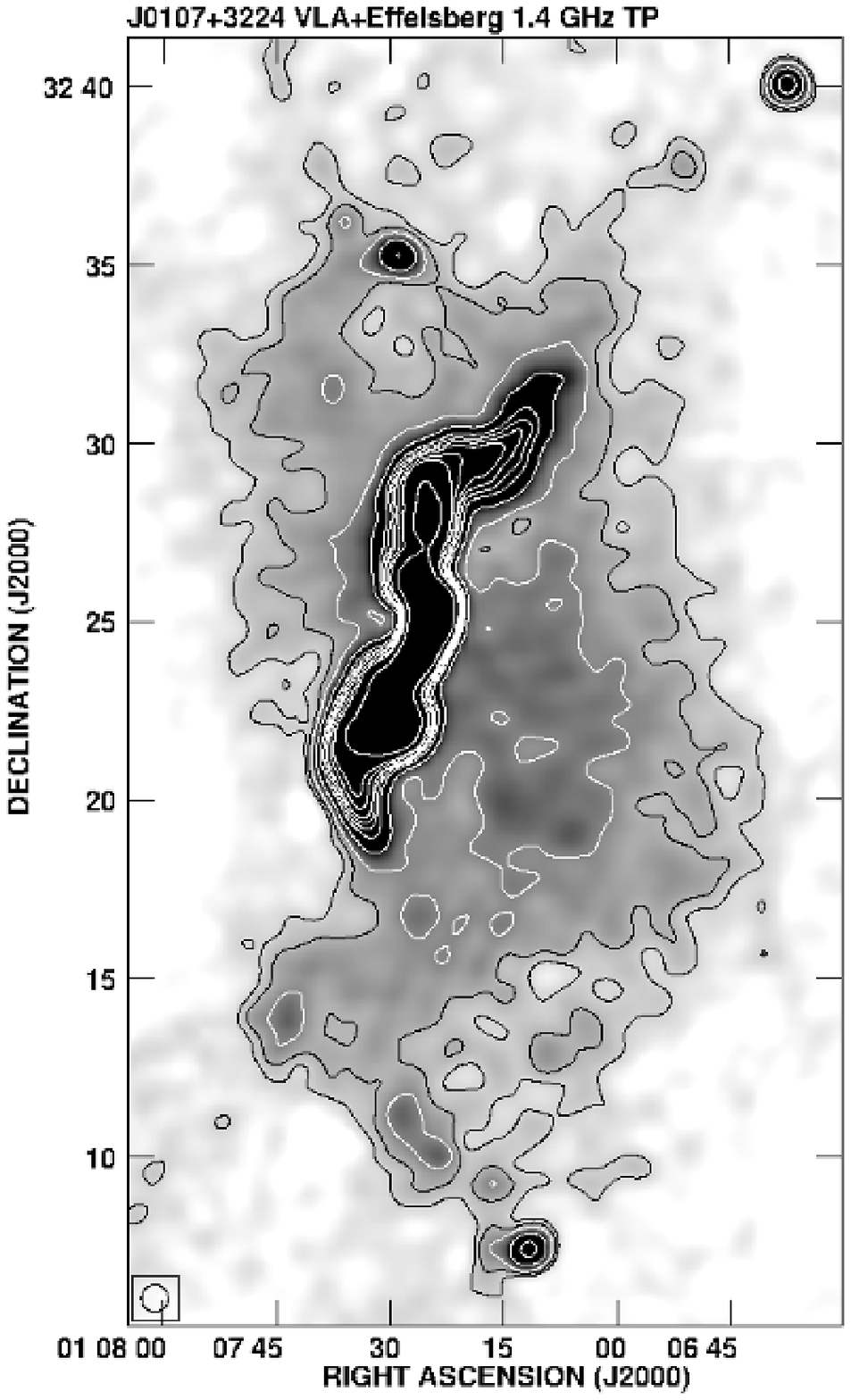}}
\FigCap{GRG J0107+3224. Noise level in both maps is at 0.6~mJy/b.a..}
\end{center}
\label{J0107}
\end{figure}

\subsection{\it GRG J0318+6829}
\label{wnb0313+683}

In the case of this source, WNB\,0313+683, the addition of the single-dish data did not enhance in any significant manner the diffuse radio emission (Fig.~3). 
The merged map shows only slightly more emission between the bright radio lobes and outside of the southern lobe. This correspondence of interferometer and single-dish
data has been noted before by~Schoenmakers et al.~(1998), who also presented a WSRT map at 325\,MHz. Its radio extent is comparable to
that of our merged map. The lack of a distinct radio cocoon around this source is reflected in the ratio of the NVSS flux to the total (merged map) flux.
As much as 70\% of the flux is visible in the NVSS map.

\begin{figure}
\begin{center}
\resizebox{0.9\hsize}{!}{\includegraphics{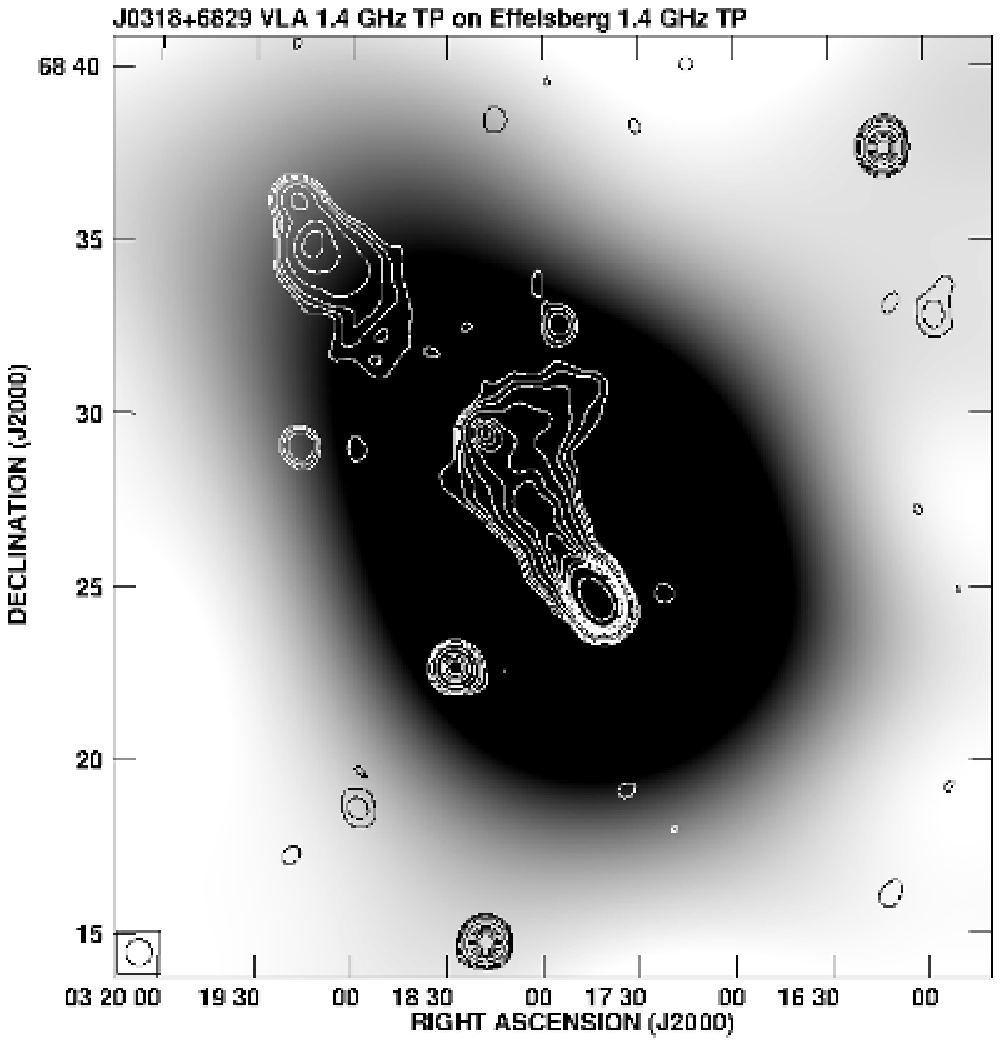}}
\resizebox{0.9\hsize}{!}{\includegraphics{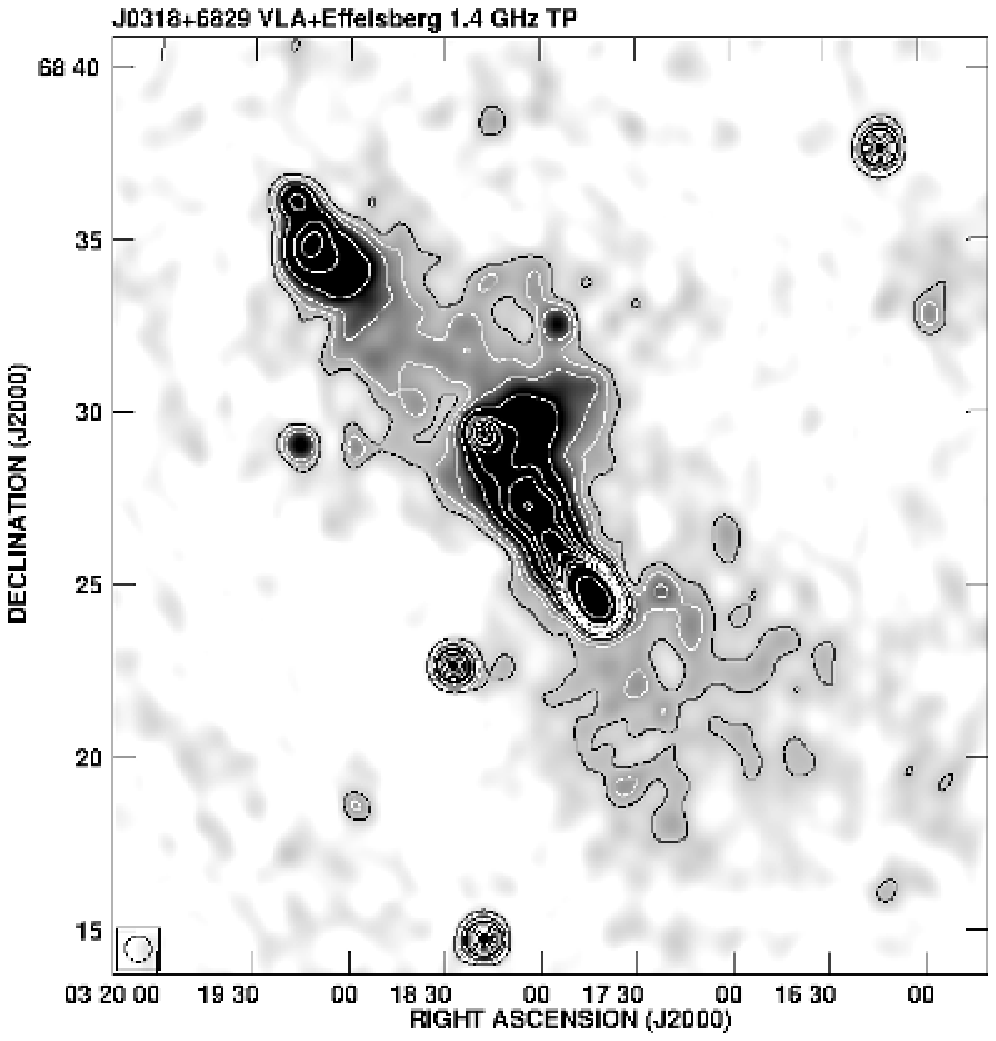}}
\FigCap{GRG J0318+6829. Noise level in both maps is at 0.5~mJy/b.a..}
\end{center}
\label{J0318}
\end{figure}

\subsection{\it GRG J0448+4502}
\label{3c129}

This source (3C\,129) forms, together with 3C\,129.1 (a compact source to the left in both panels of Fig.~4),
a galaxy cluster showing an extended X-ray halo (Taylor et al.~2001). This halo centres on 3C\,129.1 and
extends out to the ``head'' of the head-tail structure of 3C\,129. Outside of this halo, the most intense radio emission
is visible in the merged map (lower panel of Fig.~4). High-resolution radio observations at higher frequencies
performed by~Taylor et al.~(2001) revealed bright extended lobes forming the ``head'' of the source and extending further
out in the northwest direction to form the ``tail'' structure visible in the NVSS image (upper panel of Fig.~4).
The merged map shows a radio halo in the position of the source, truncated on its northern side. It is likely that the morphology
of the source (thus also its halo as well) is significantly affected by some movement in the cluster's periphery, just outside of the
X-ray halo of the cluster. However, the sharp edge of the radio emission visible in the east is likely caused
by the NVSS map showing an artifact introduced by the {\it CLEAN}ing. This artifact includes an area of significantly negative
flux that could not be fully filled in the merging process. Still, this truncation is found at the outer edge
of the bright X-ray halo~(Taylor et al.~2001), so it is possible that we observe a combination of both effects.
We also note that the visible radio emission south of 3C\,129 is likely a part of the cluster's radio halo rather than a cocoon around
this source. Consequently, we did not include this emission in the total flux from 3C\,129.
Lane et al.~(2002) presented a 90\,cm map of this source, which suggested an existence
of a fossil radio plume in this system. A radio bridge with 3C\,129.1 can also be seen.
For this source, it was possible to recover in total 18\% of flux from the single-dish data.

\begin{figure}
\begin{center}
\resizebox{\hsize}{!}{\includegraphics[angle=-90]{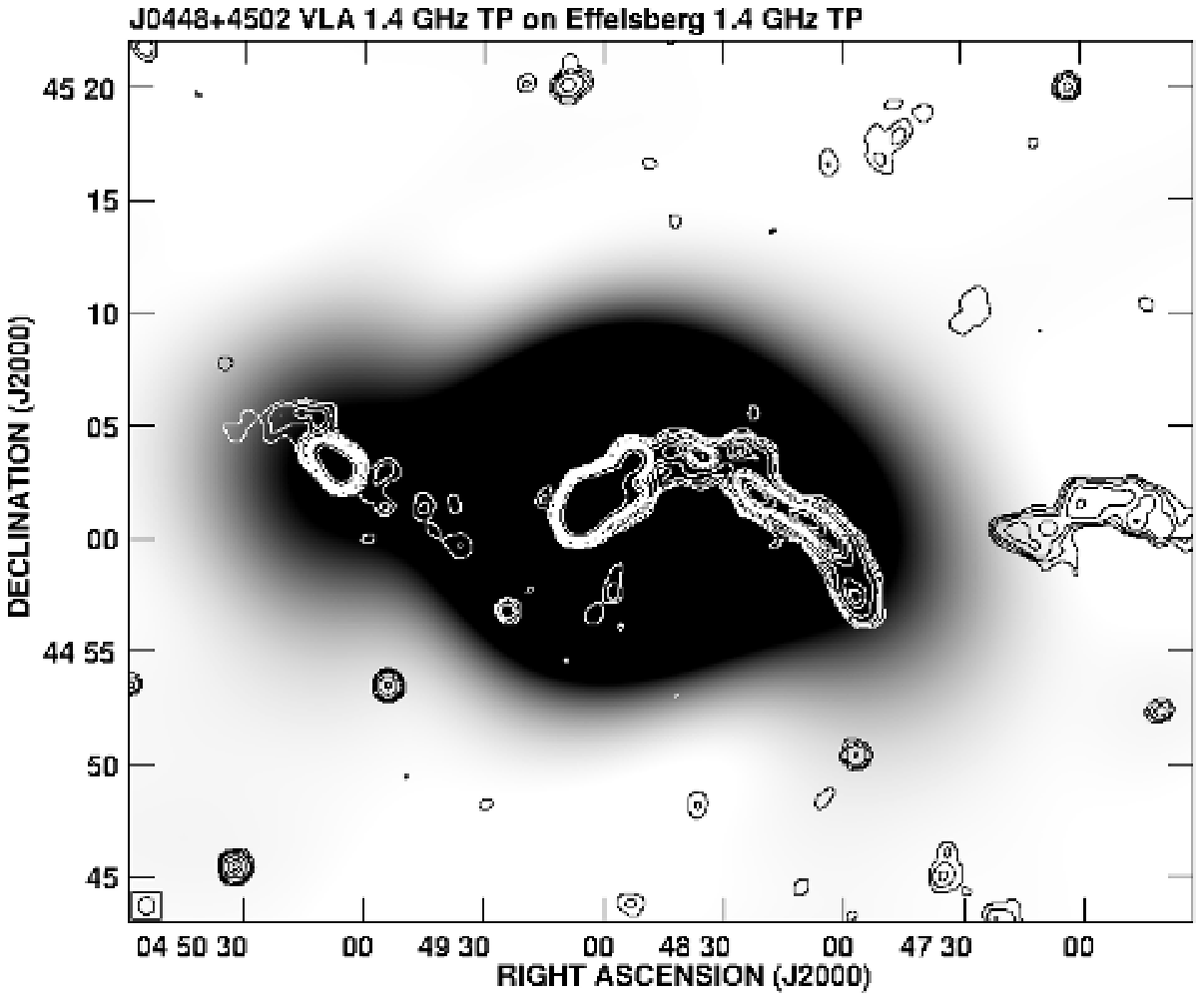}}
\resizebox{\hsize}{!}{\includegraphics[angle=-90]{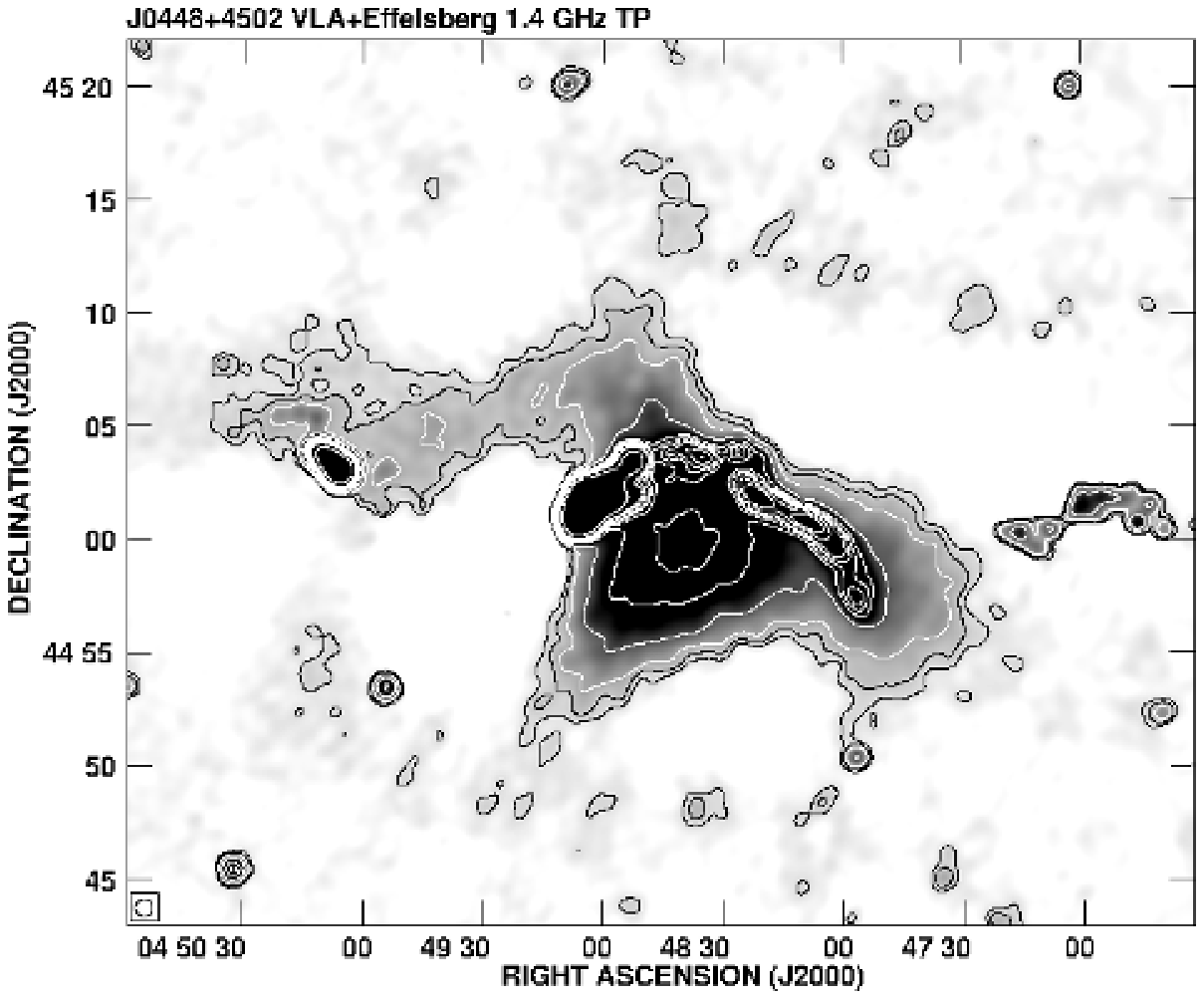}}
\FigCap{GRG J0448+4502. Noise level in both maps is at 0.6~mJy/b.a..}
\end{center}
\label{J0448}
\end{figure}

\subsection{\it GRG J0702+4859}
\label{grg0702}

This radio source lacks any signs of a radio halo (cocoon), as the merged map looks almost exactly like the NVSS one (see both panels of Fig.~5). 
The only visible difference is a small extension south of the western lobe. Nevertheless, this extension is at a 3$\sigma$ level only. What is a bit
surprising, the NVSS map includes only 77\% of the total (merged map) flux. This will be further discussed in Sect.~4.

\begin{figure}
\begin{center}
\resizebox{\hsize}{!}{\includegraphics{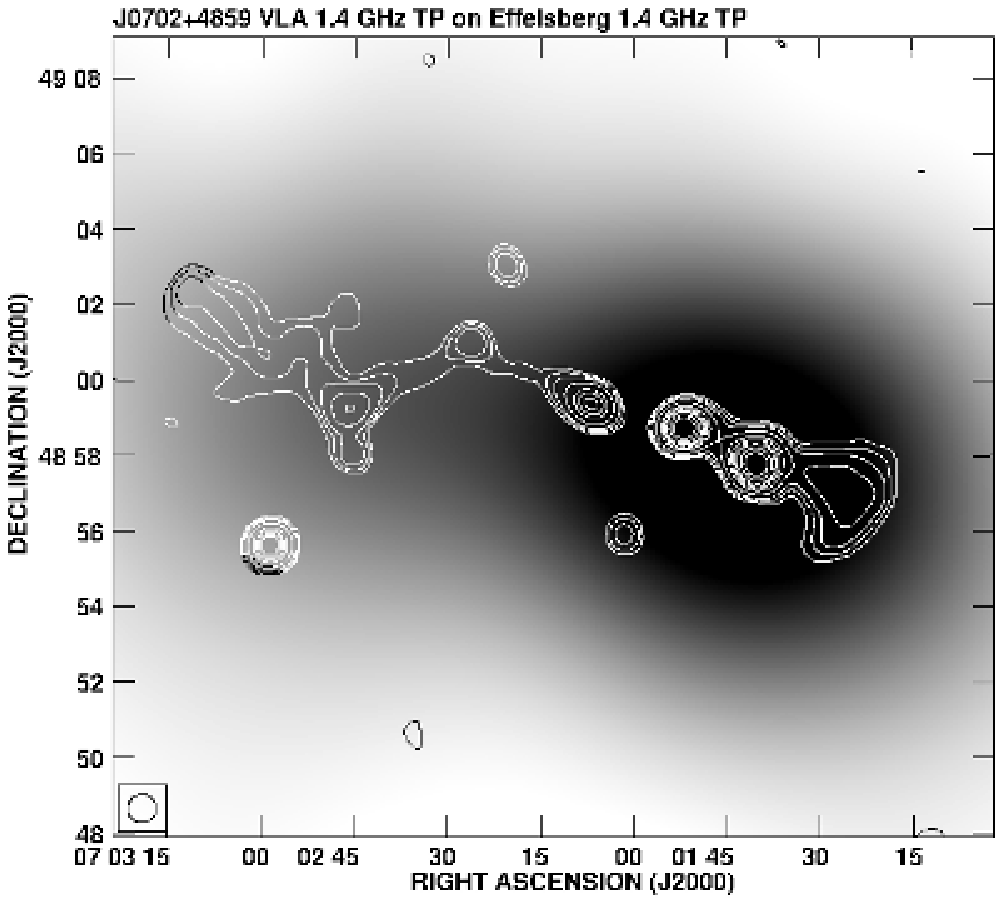}}
\resizebox{\hsize}{!}{\includegraphics{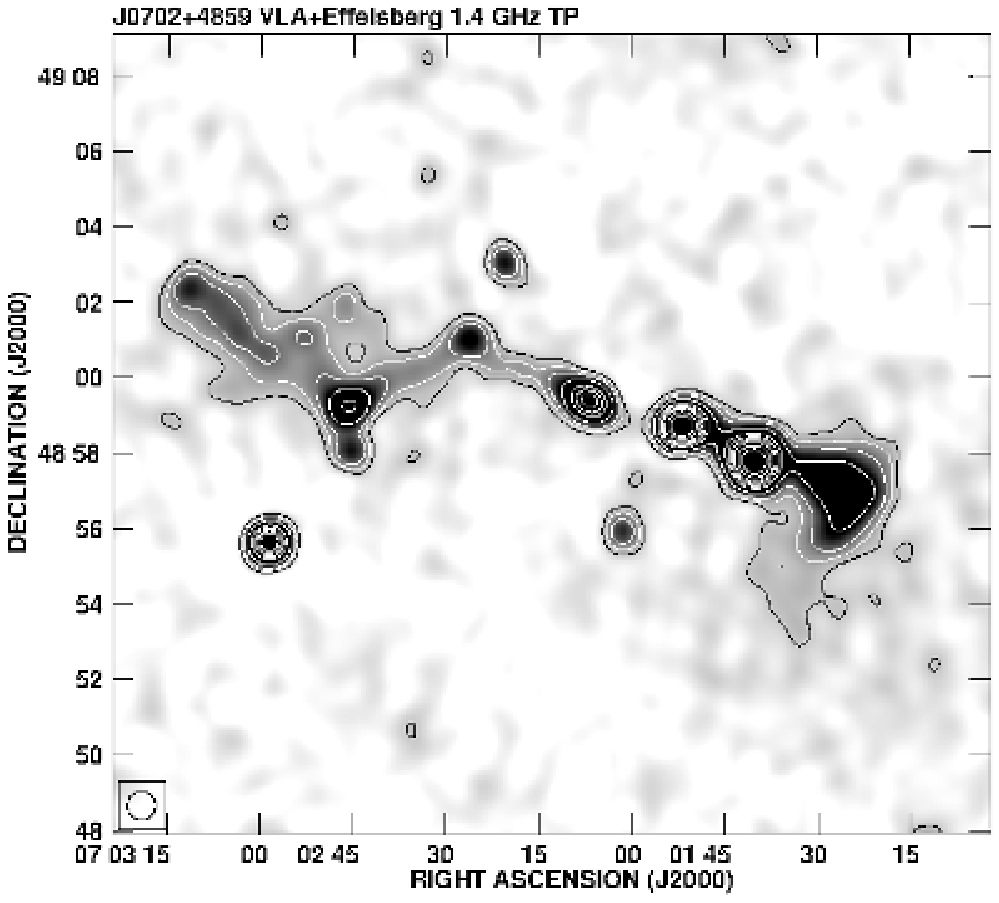}}
\FigCap{GRG J0702+4859. Noise level in both maps is at 0.5~mJy/b.a..}
\end{center}
\label{J0702}
\end{figure}

\subsection{\it GRG J0748+5548}
\label{da240}

Similarly to NGC\,315 and NGC\,6251, the merged map of DA\,240 (lower panel of Fig.~6) shows the extent and morphology of the radio emission
similar to that visible in the 609\,MHz WSRT map shown by~Mack et al.~(1997). Here again, as in 3C\,31 and also in 4C\,73.08,
a multiple jet activity can be suggested.

\begin{figure}
\begin{center}
\resizebox{\hsize}{!}{\includegraphics{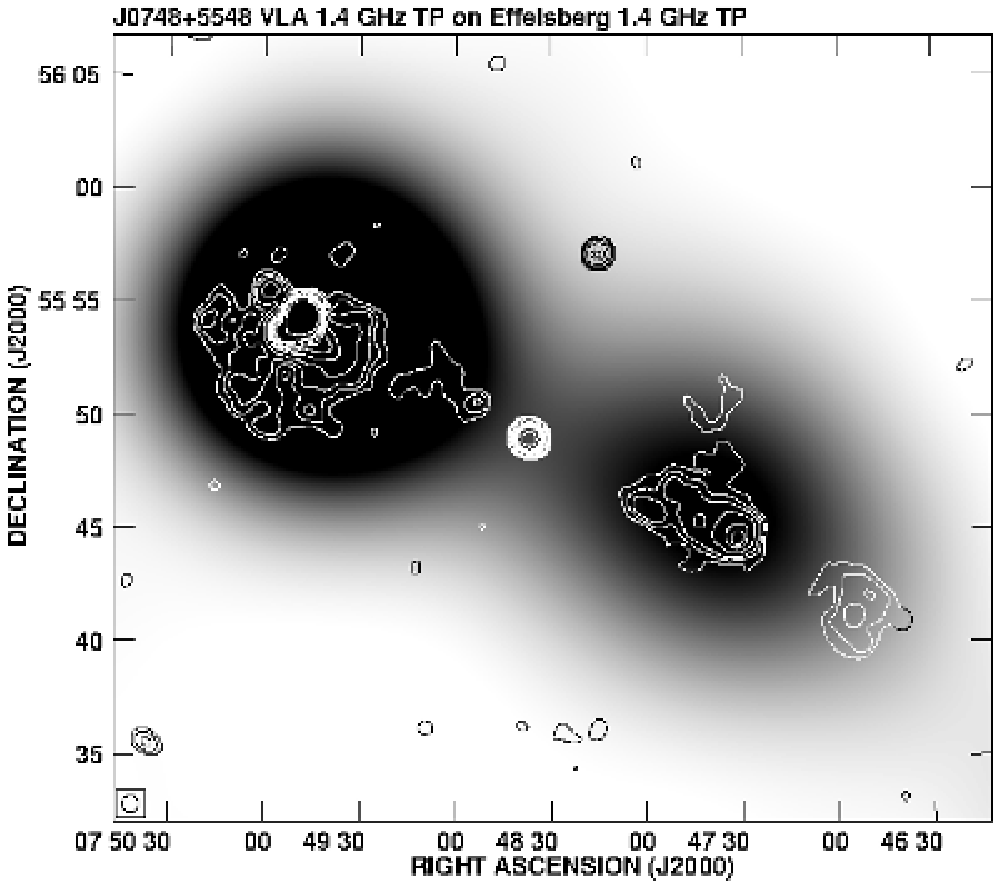}}
\resizebox{\hsize}{!}{\includegraphics{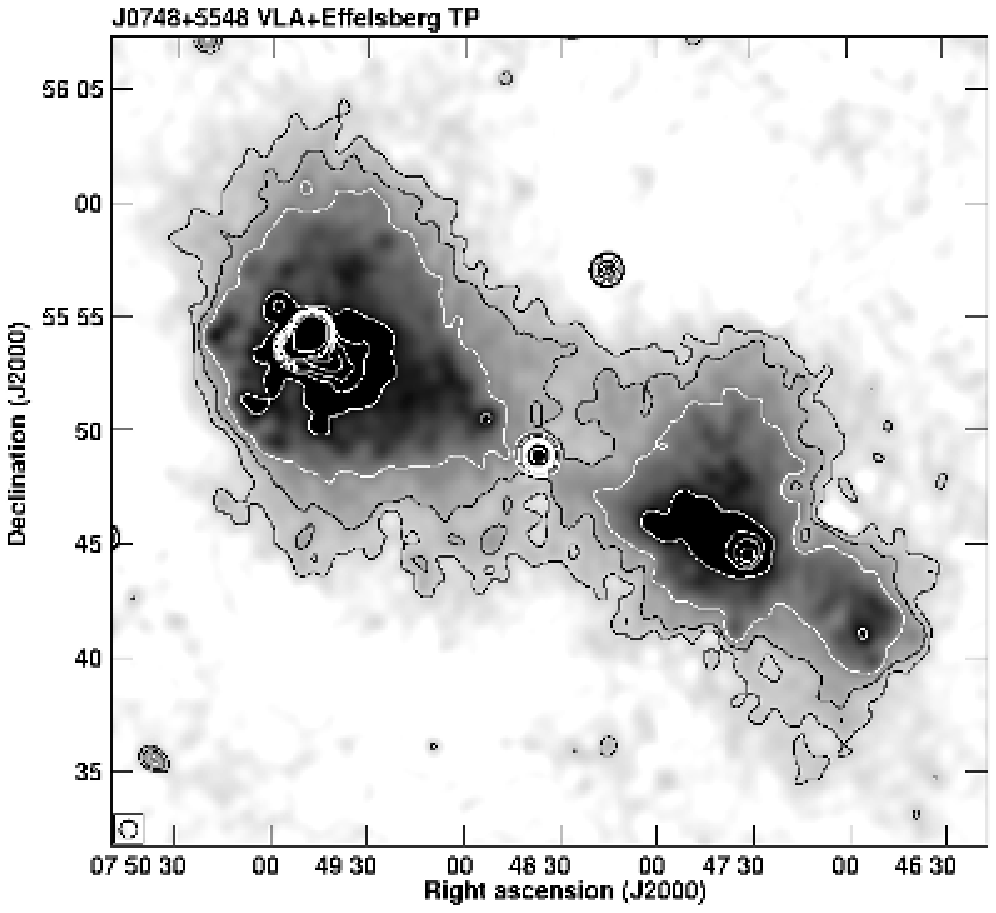}}
\FigCap{GRG J0748+5548. Noise level in both maps is at 0.6~mJy/b.a..}
\end{center}
\label{J0748}
\end{figure}

\subsection{\it GRG J0949+7314}
\label{4C73.08}

The radio galaxy 4C\,73.08 was first studied by~Mayer~(1979), who noted that
its luminosity is in the boundary region between FRI and FRII class sources.
Also its X-shaped morphology does not naturally fit into either FR
class. Basing on the 327, 608.5, and 1412\,MHz radio total intensity and linear
polarization maps~J\"agers~(1986) concluded that the peculiar structure of
this source can be explained as a distortion of a linear radio galaxy due to 
the motion of the parent galaxy combined with the precessional motion of the
nucleus. Strom et al.~(2013) observed this source with the Westerbork
Synthesis Radio Telescope at 1381.1\,MHz. In their deep map,
a double-lobed structure can be clearly seen, with broad components that are
nearly as wide as long. The hotspot in the northern lobe is more
centrally located and considerably fainter than that in the southern one. There is
a small but significant plateau of fainter emission to the west of the
southern hotspot. Therefore, both lobes of 4C\,73.08 have a forward
protrusion of emission that pushes ahead the leading edge of the diffuse
lobes. Furthermore, as shown by~J\"agers~(1986), the southern lobe 
revealed an intriguing distribution of spectral indices which was quite
different from that of classical double sources. There is a
tendency for the spectrum to steepen towards the lateral edges of the lobe.
This resembles the morphology of the relic radio emission seen in 3C\,388~(Roettiger et al.~1994), 
in which multiple epochs of activity are to be seen.
J\"agers~(1986) suggested also that the visible hot spots are more recent
elements of enhanced jet activity, immersed in the more diffuse older
structure. Our merged map (lower panel of Fig.~7) seems to closely confirm such predictions, as it shows
a larger extent of the diffuse radio emission with the bright lobes, deeply embedded
in the outer structures.

\begin{figure}
\begin{center}
\resizebox{0.9\hsize}{!}{\includegraphics{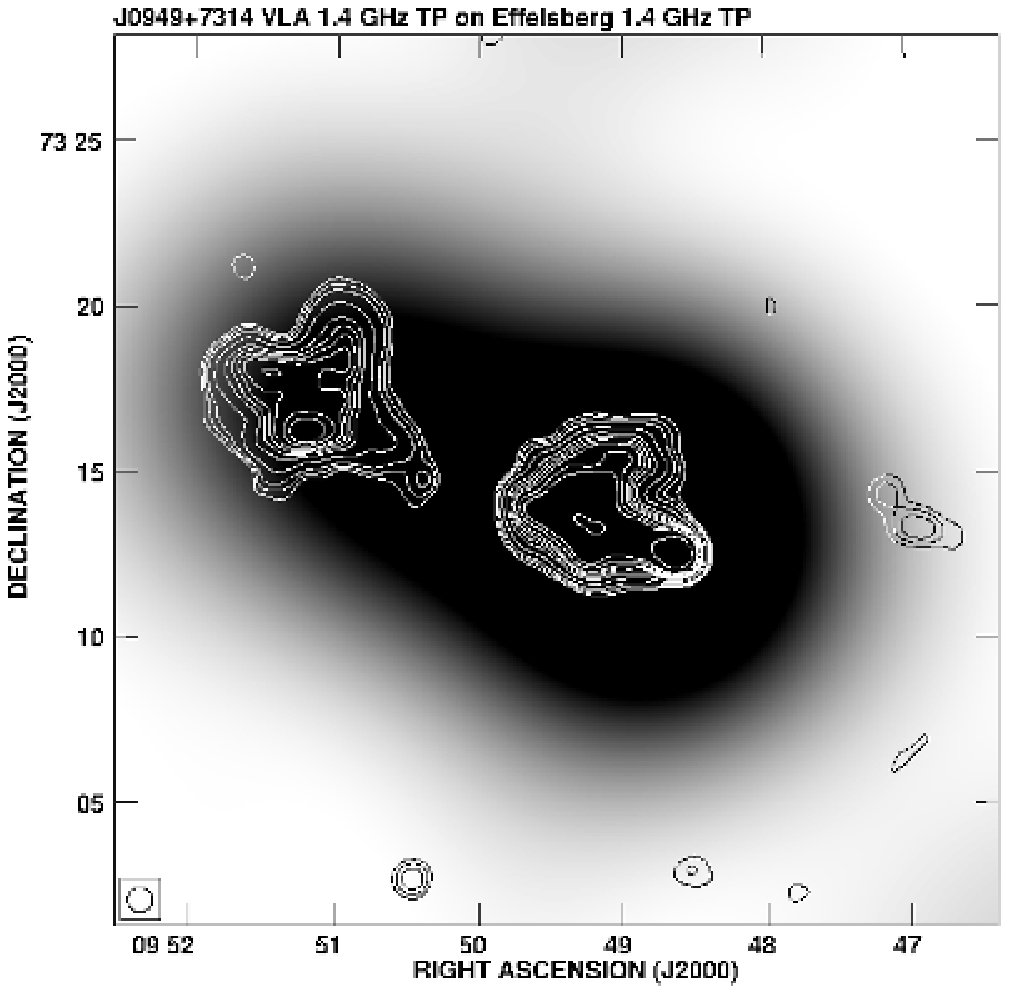}}
\resizebox{0.9\hsize}{!}{\includegraphics{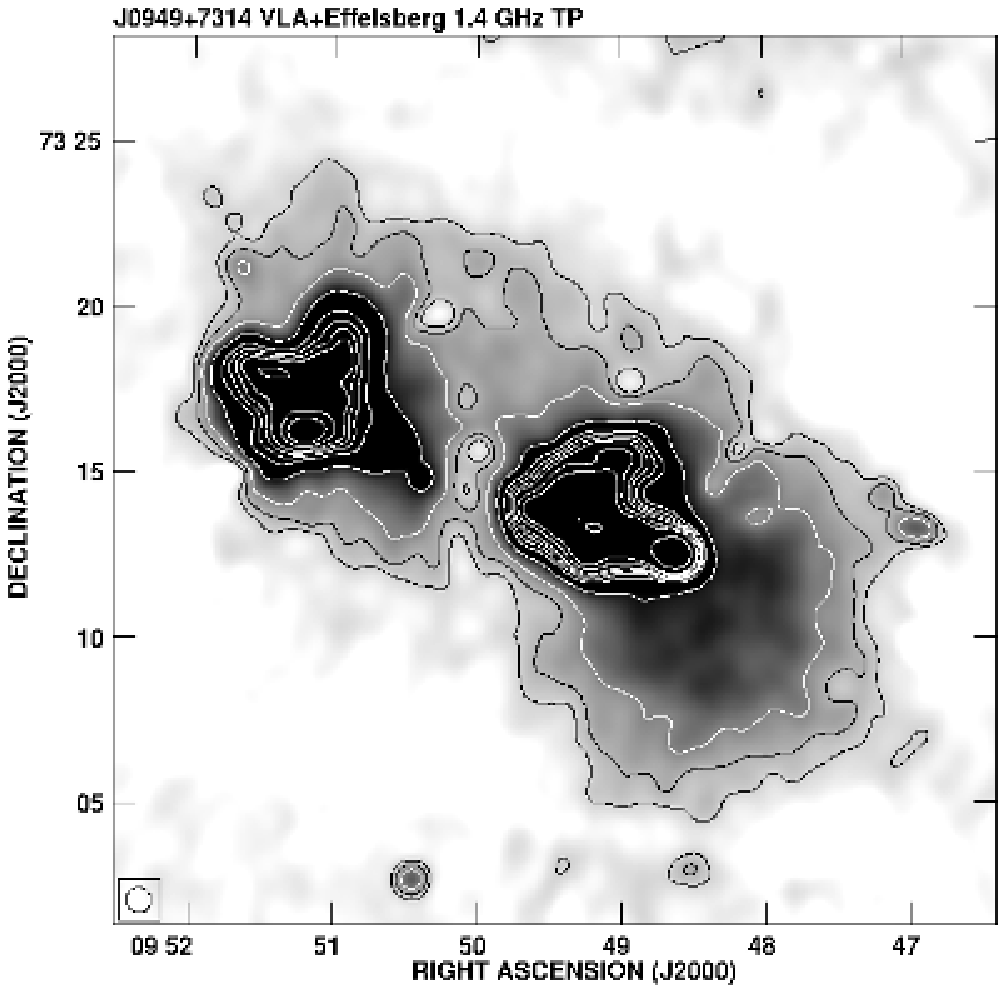}}
\FigCap{GRG J0949+7314. Noise level in both maps is at 0.6~mJy/b.a..}
\end{center}
\label{J0949}
\end{figure}

\subsection{\it GRG J1006+3454}
\label{3c236}

In general, 3C\,236 does not show any evidence of an extended diffuse radio halo, except for the
emission enhancement outside its central source (Fig.~8, bottom). Although it looks like an
artifact introduced by the merging process and the very strong central source,
a slight extension towards this position is also visible in the 325\,MHz WSRT map by~Mack et al.~(1997).
The NVSS map (contours in the upper panel of Fig.~8)
shows 82\% of the total flux of the merged map.

\begin{figure}
\begin{center}
\resizebox{\hsize}{!}{\includegraphics{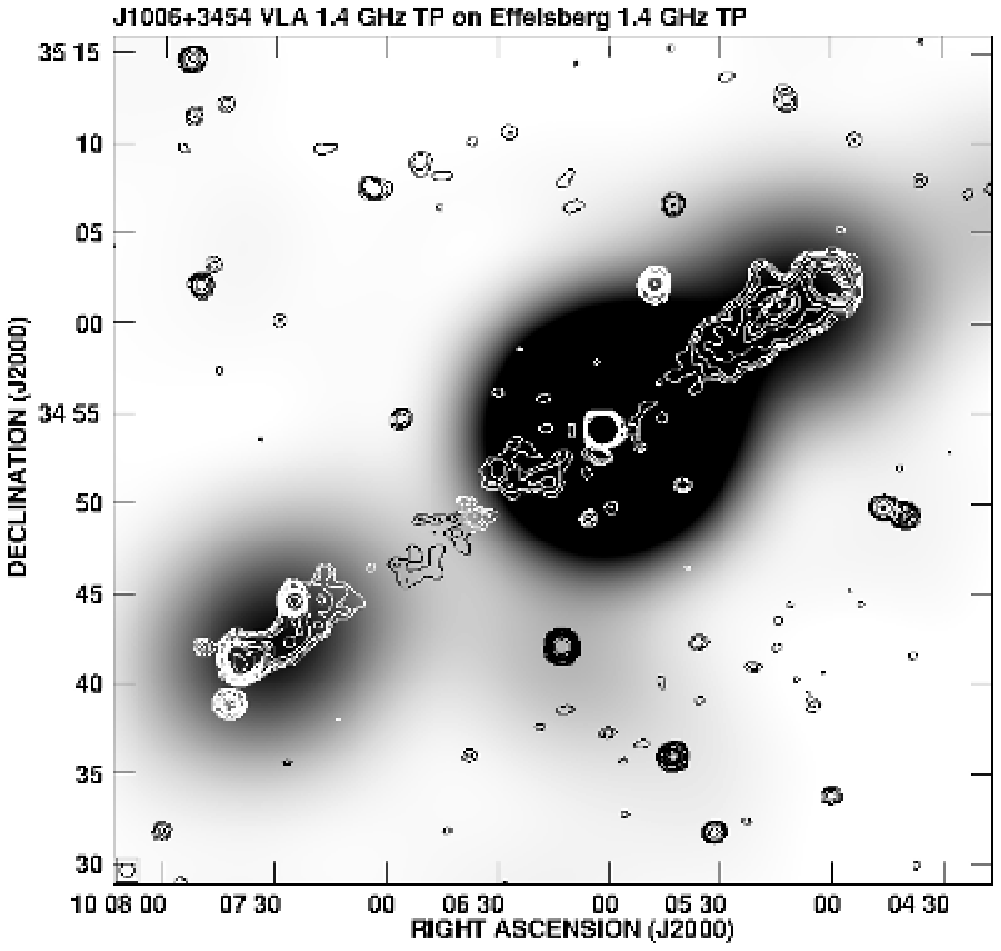}}
\resizebox{\hsize}{!}{\includegraphics{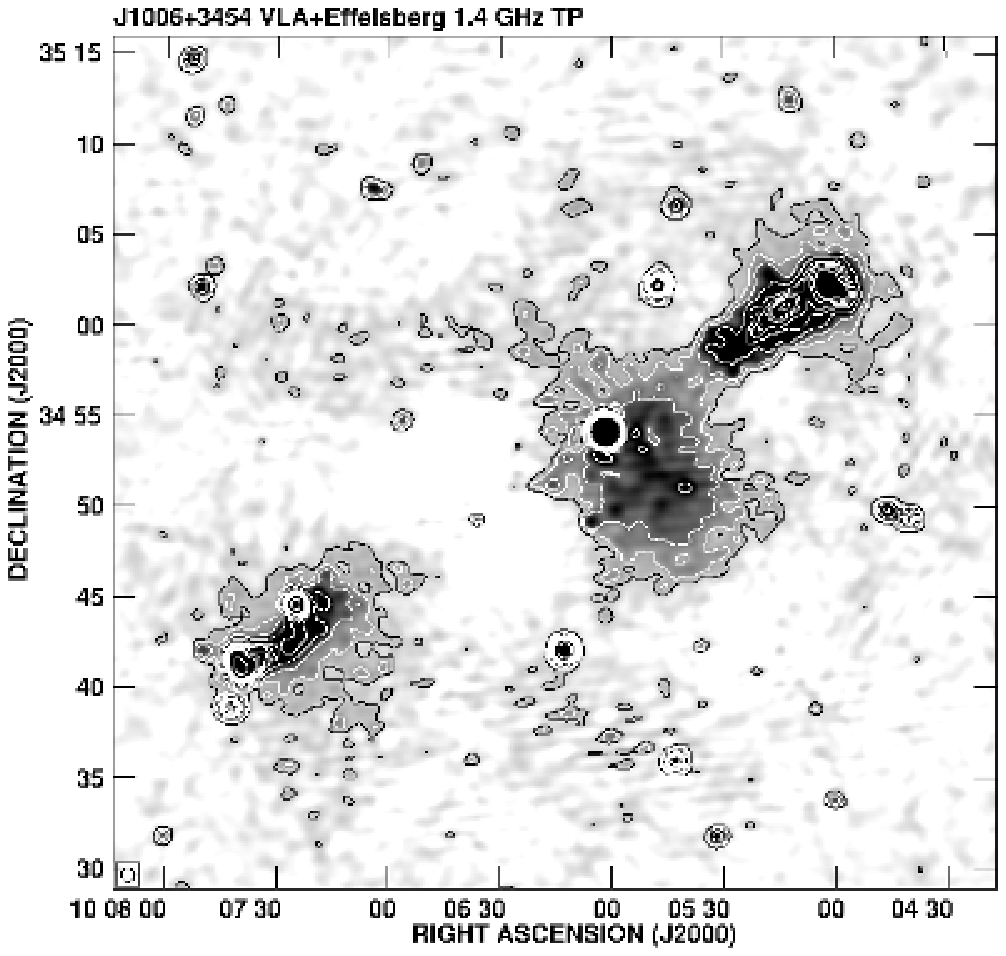}}
\FigCap{GRG J1006+3454. Noise level in both maps is at 0.5~mJy/b.a..}
\end{center}
\label{J1006}
\end{figure}

\subsection{\it GRG J1032+5644}
\label{hb13}

A ``patchy'' structure outside the main axis of this source (Fig.~9, bottom) suggests that an extended radio halo
might exist at lower frequencies.
Especially as despite a similar extent of the radio source, both in the NVSS and the merged map,
the former map includes only 69\% of the latter map's flux.

\begin{figure}
\begin{center}
\resizebox{0.7\hsize}{!}{\includegraphics{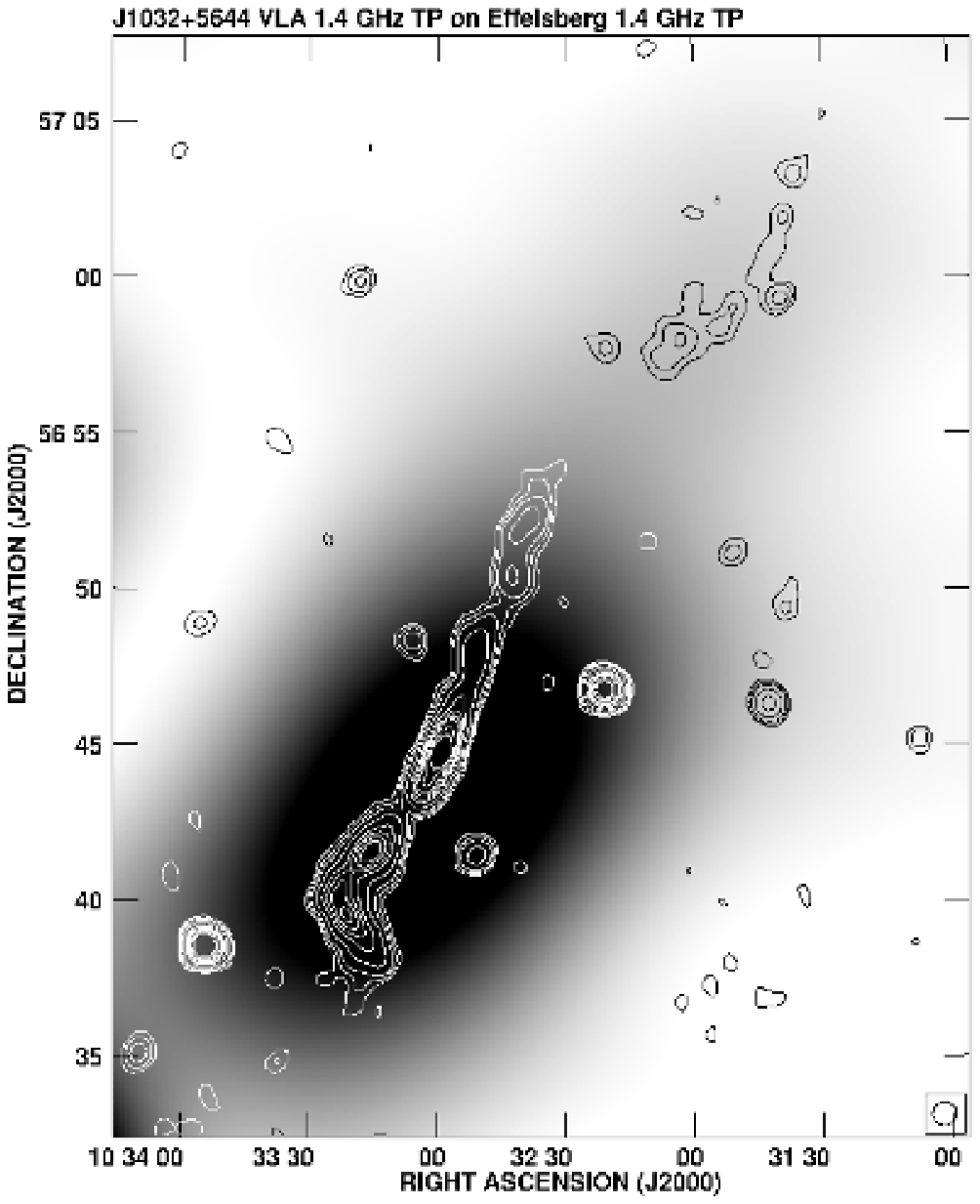}}
\resizebox{0.7\hsize}{!}{\includegraphics{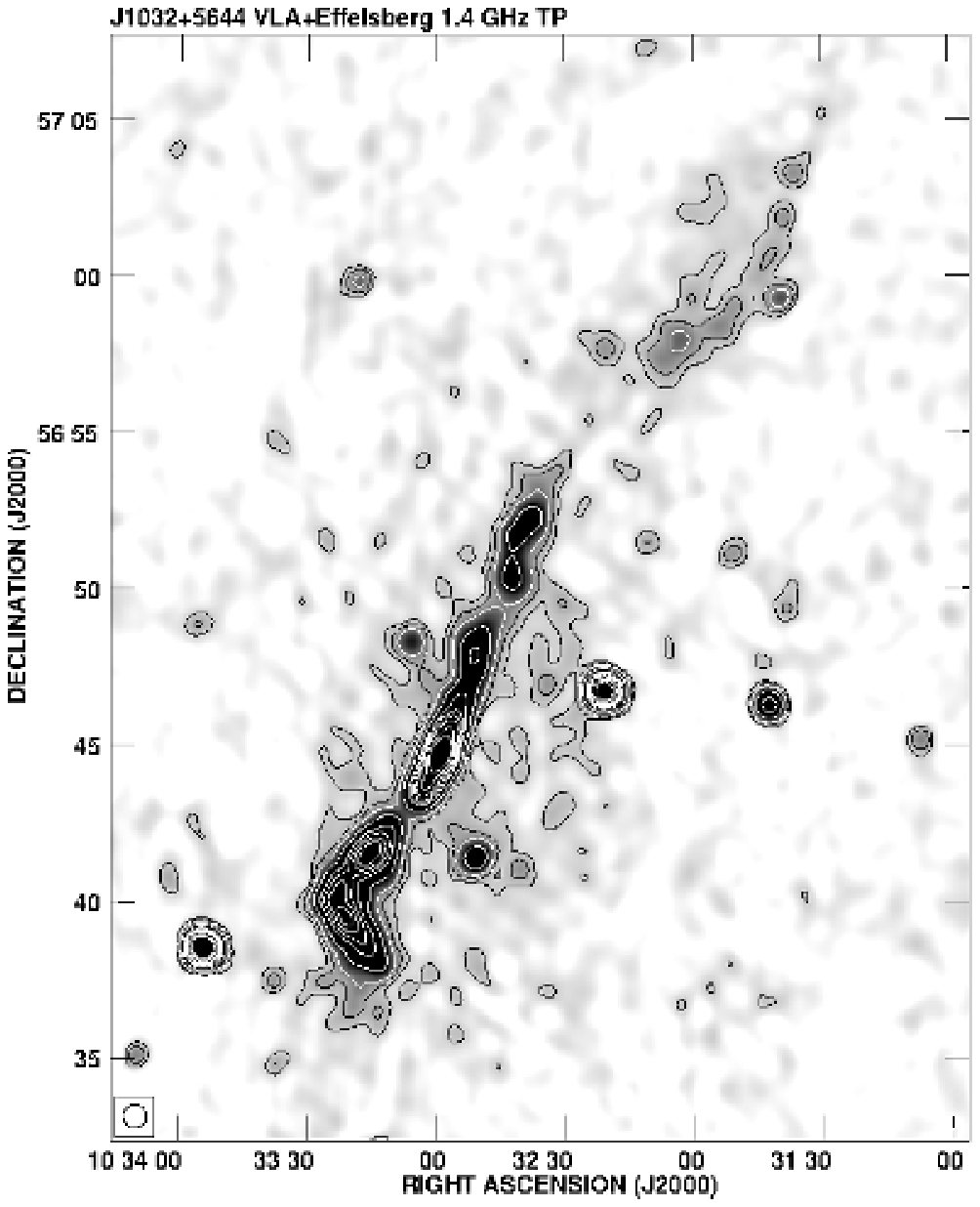}}
\end{center}
\FigCap{GRG J1032+5644. Noise level in both maps is at 0.5~mJy/b.a..}
\label{J1032}
\end{figure}

\subsection{\it GRG J1312+4450}
\label{grg1312}

A better sensitivity to extended structures of the merged map (Fig.~10, bottom) allows to trace both narrow
jets to a greater distance from the central source. Both lobes, however, do not show any diffuse extended emission.
Either in the NVSS or in the merged maps only the hot spots are visible.

\begin{figure}
\begin{center}
\resizebox{\hsize}{!}{\includegraphics[angle=-90]{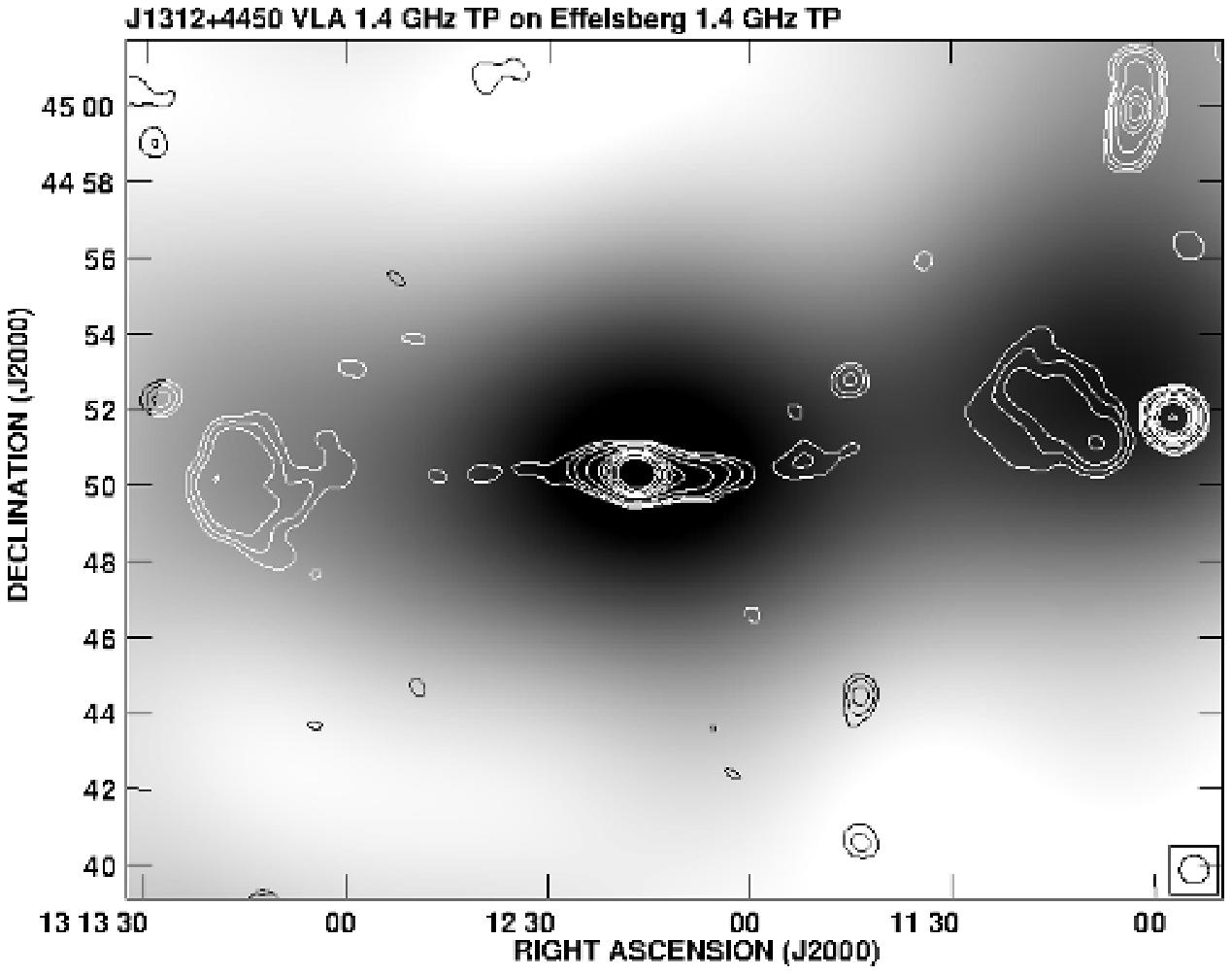}}
\resizebox{\hsize}{!}{\includegraphics[angle=-90]{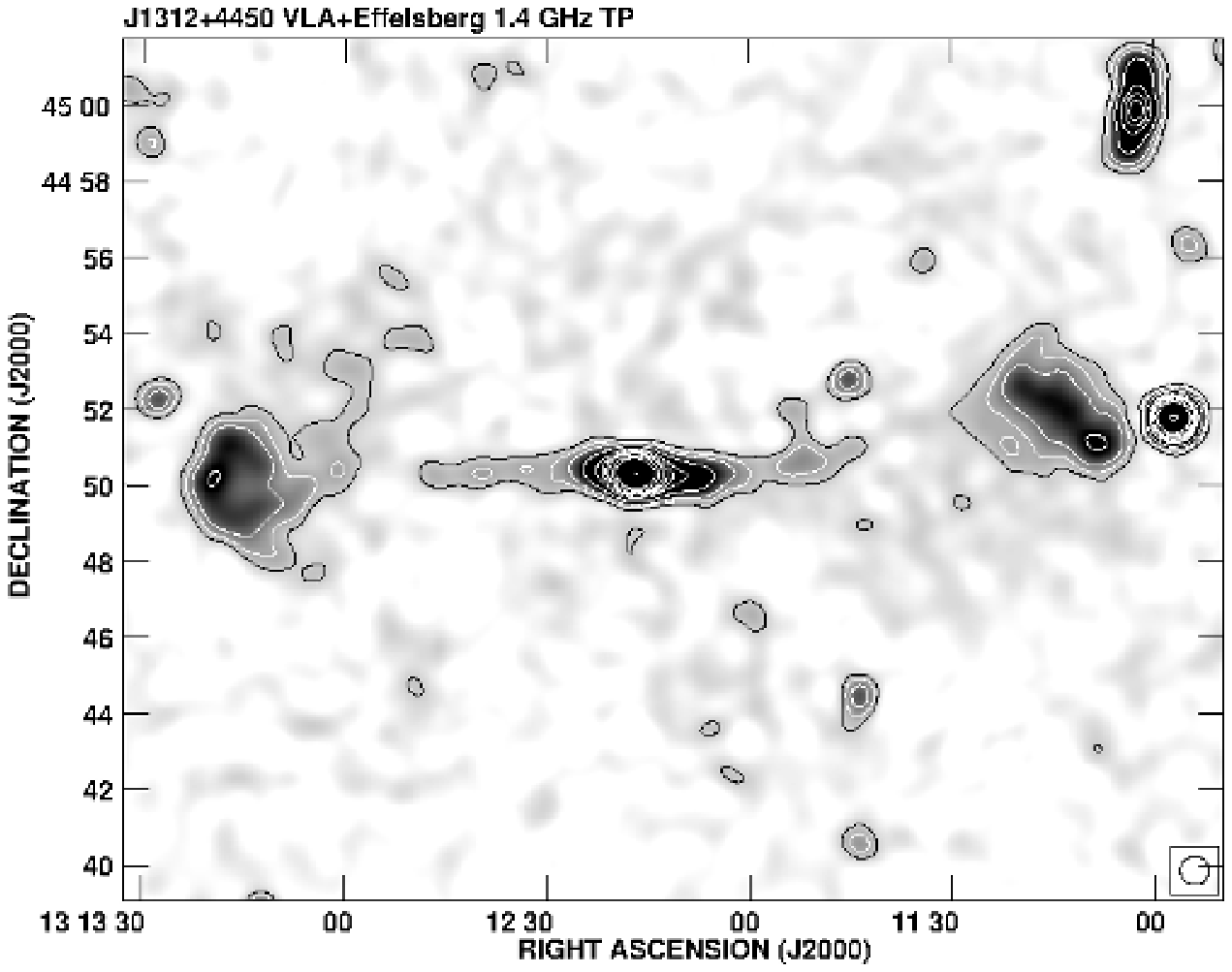}}
\FigCap{GRG J1312+4450. Noise level in both maps is at 0.5~mJy/b.a..}
\end{center}
\label{J1312}
\end{figure}

\subsection{\it GRG J1428+2918}
\label{grg1428}

This source shows no signs of any diffuse radio halo (Fig.~11). However, the merged map reveals continuous emission
between the central source and both lobes. The NVSS map includes 82\% of the total (merged map) flux.

\begin{figure}
\begin{center}
\resizebox{\hsize}{!}{\includegraphics{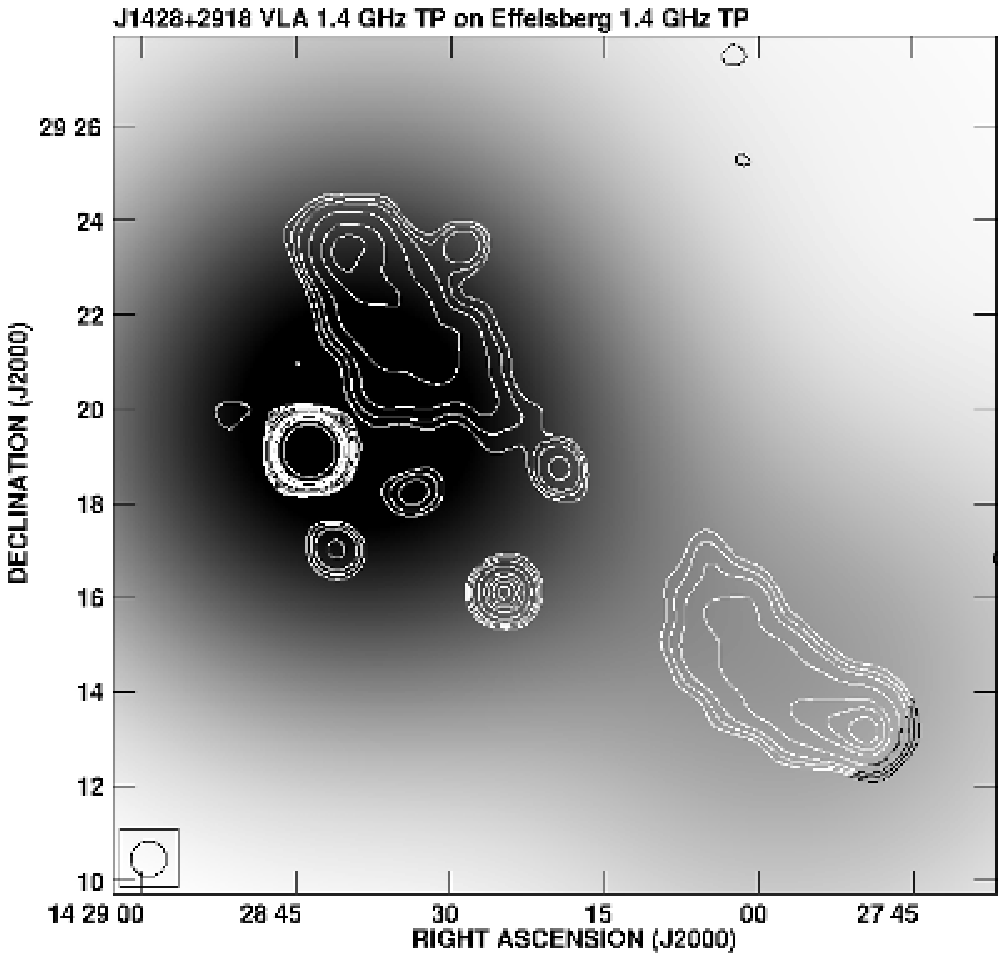}}
\resizebox{\hsize}{!}{\includegraphics{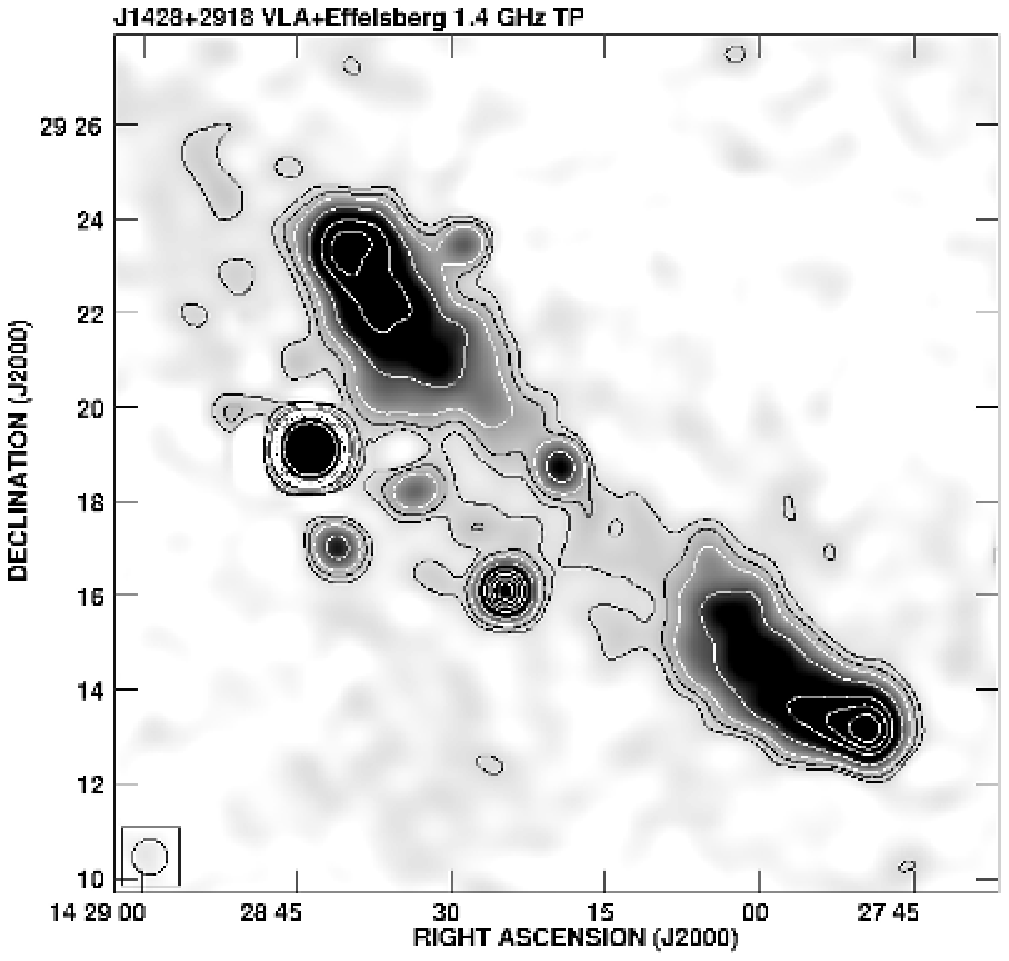}}
\FigCap{GRG J1428+2918. Noise level in both maps is at 0.5~mJy/b.a..}
\end{center}
\label{J1428}
\end{figure}

\subsection{\it GRG J1552+2005}
\label{3c326}

An extended diffuse radio halo around 3C\,326 (Fig.~12) suggests an earlier period of activity in the past with ``hot spots''
visible in its centre to mark the present radio activity. This halo causes a significant depletion in the radio flux of the NVSS map,
which shows only 54\% of the total flux. Possibly due to its low surface-brightness, the halo is not visible at lower frequencies, including
the maps at 325\,MHz and 609\,MHz presented by~Mack et al.~(1997).

\begin{figure}
\begin{center}
\resizebox{\hsize}{!}{\includegraphics[angle=-90]{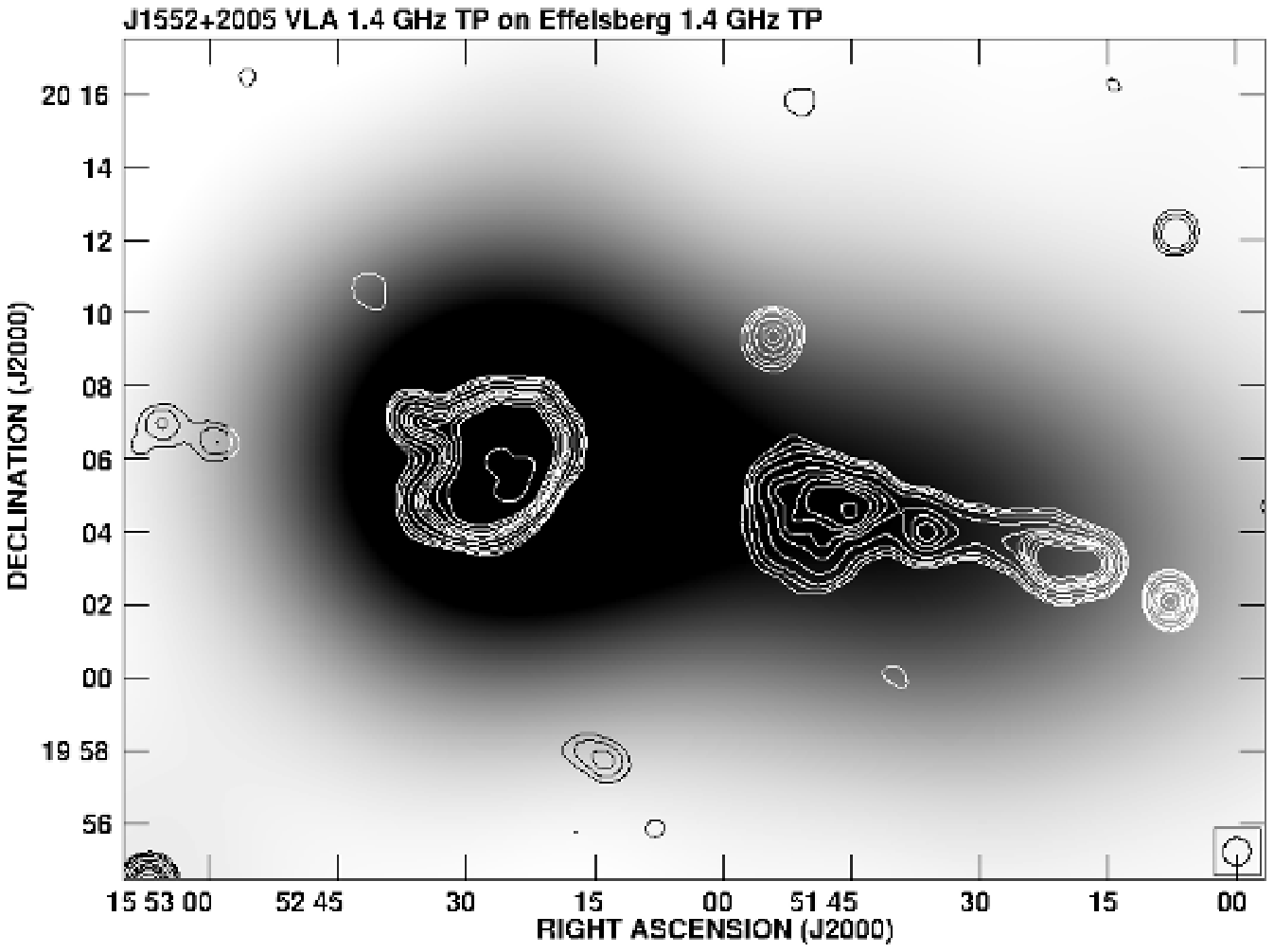}}
\resizebox{\hsize}{!}{\includegraphics[angle=-90]{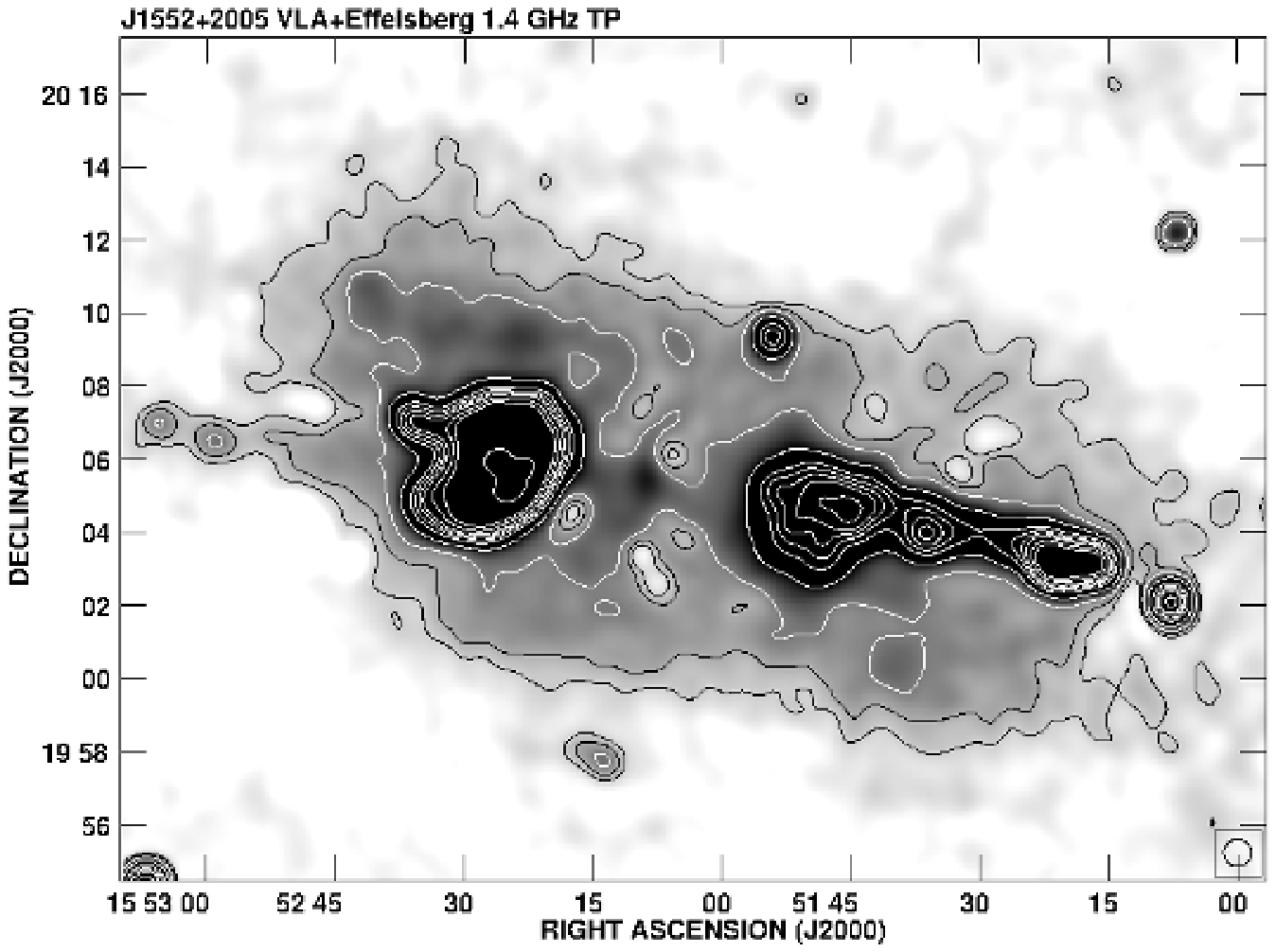}}
\FigCap{GRG J1552+2005. Noise level in both maps is at 0.6~mJy/b.a..}
\end{center}
\label{J1552}
\end{figure}

\subsection{\it GRG J1628+5146}
\label{mrk1498}

This radio galaxy is another source showing no clues for any radio halo. The single-dish data contributed only
with diffuse emission that connects the central source with the northern lobe (Fig.~13, bottom). Also the lower frequency WENSS data
presented by~R\"ottgering et al.~(1996) show no signs of diffuse radio emission.

\begin{figure}
\begin{center}
\resizebox{0.5\vsize}{!}{\includegraphics{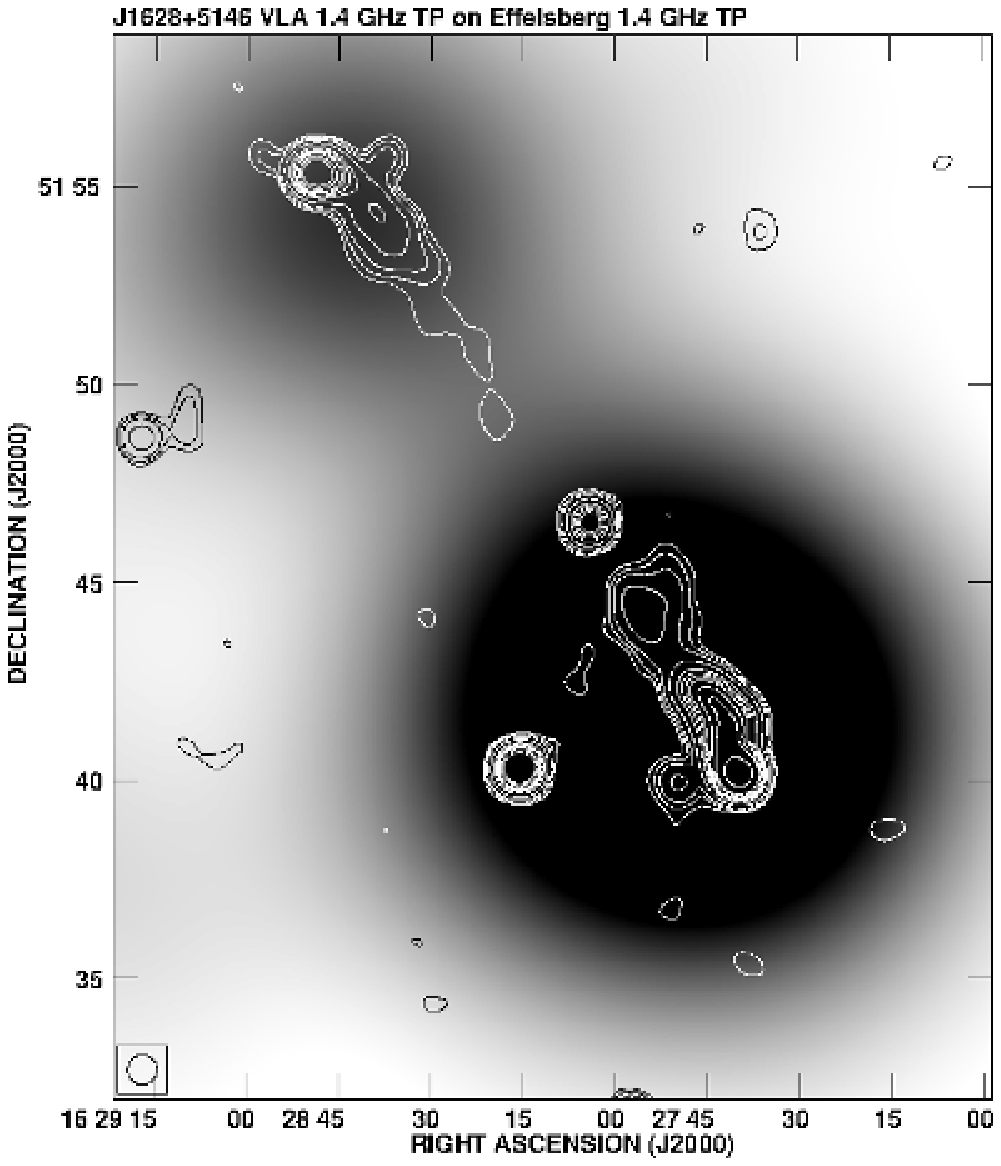}}
\resizebox{0.5\vsize}{!}{\includegraphics{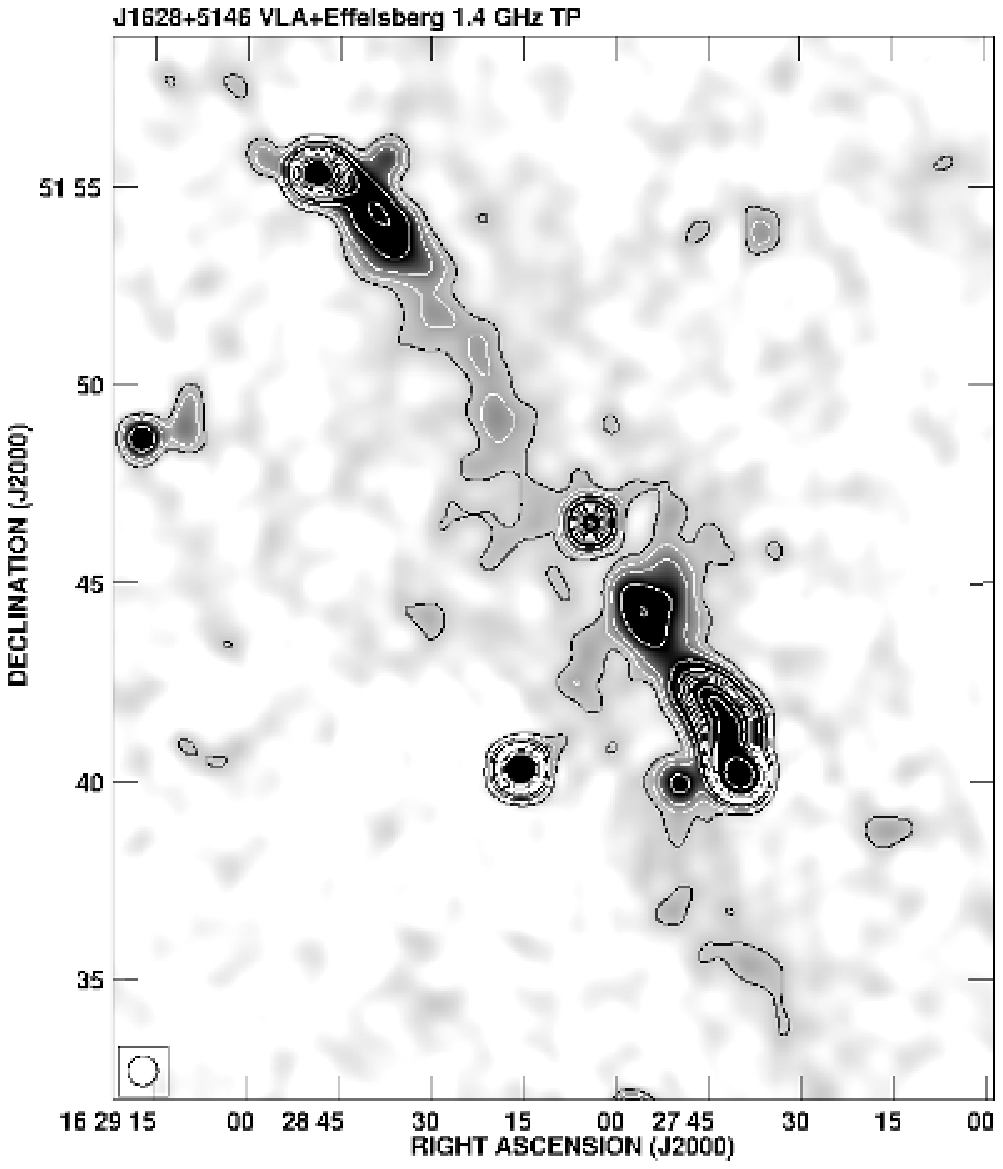}}
\FigCap{GRG J1628+5146. Noise level in both maps is at 0.5~mJy/b.a..}
\end{center}
\label{J1628}
\end{figure}

\subsection{\it GRG J1632+8232}
\label{ngc6251}

The overall morphology of NGC\,6251 at 1.4\,GHz (lower panel of Fig.~14) resembles that observed by~Mack et al.~(1997) with the WSRT at much lower frequency of 325\,MHz. The NVSS data alone
suggest that we see a double-double structure produced by this radio galaxy. A more extended diffuse radio
lobe in the eastern side of the merged map and coaxiality of both the outer and inner structures could support such
a scenario. Especially, that the outer structures seem to have steeper spectral indices, as visible in the multifrequency
maps shown by~Mack et al.~(1997). An inspection of the NVSS data showed that 51\% of the flux has been
lost due to interferometric observations when compared with the merged map.

\begin{figure}
\begin{center}
\resizebox{\hsize}{!}{\includegraphics[angle=-90]{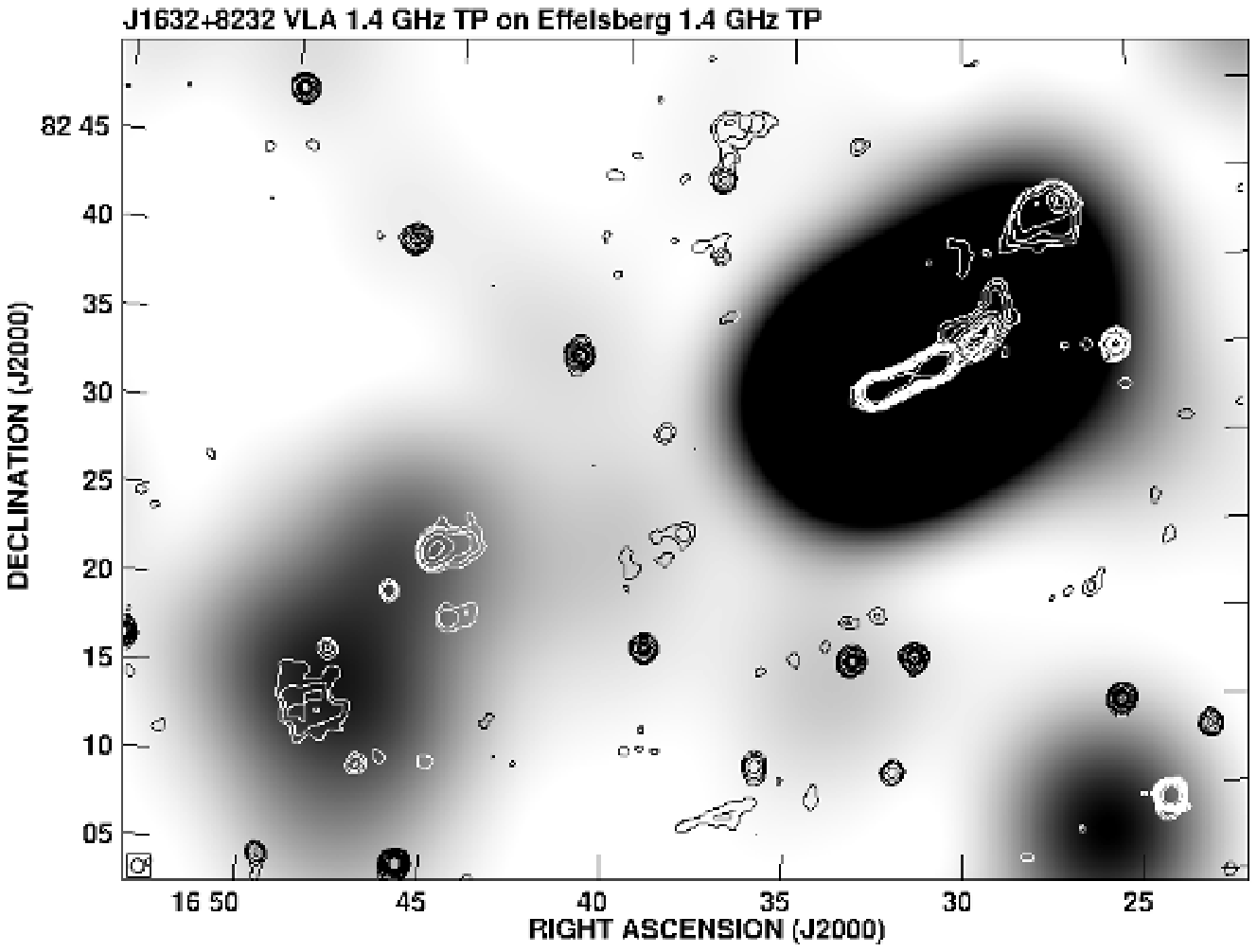}}
\resizebox{\hsize}{!}{\includegraphics[angle=-90]{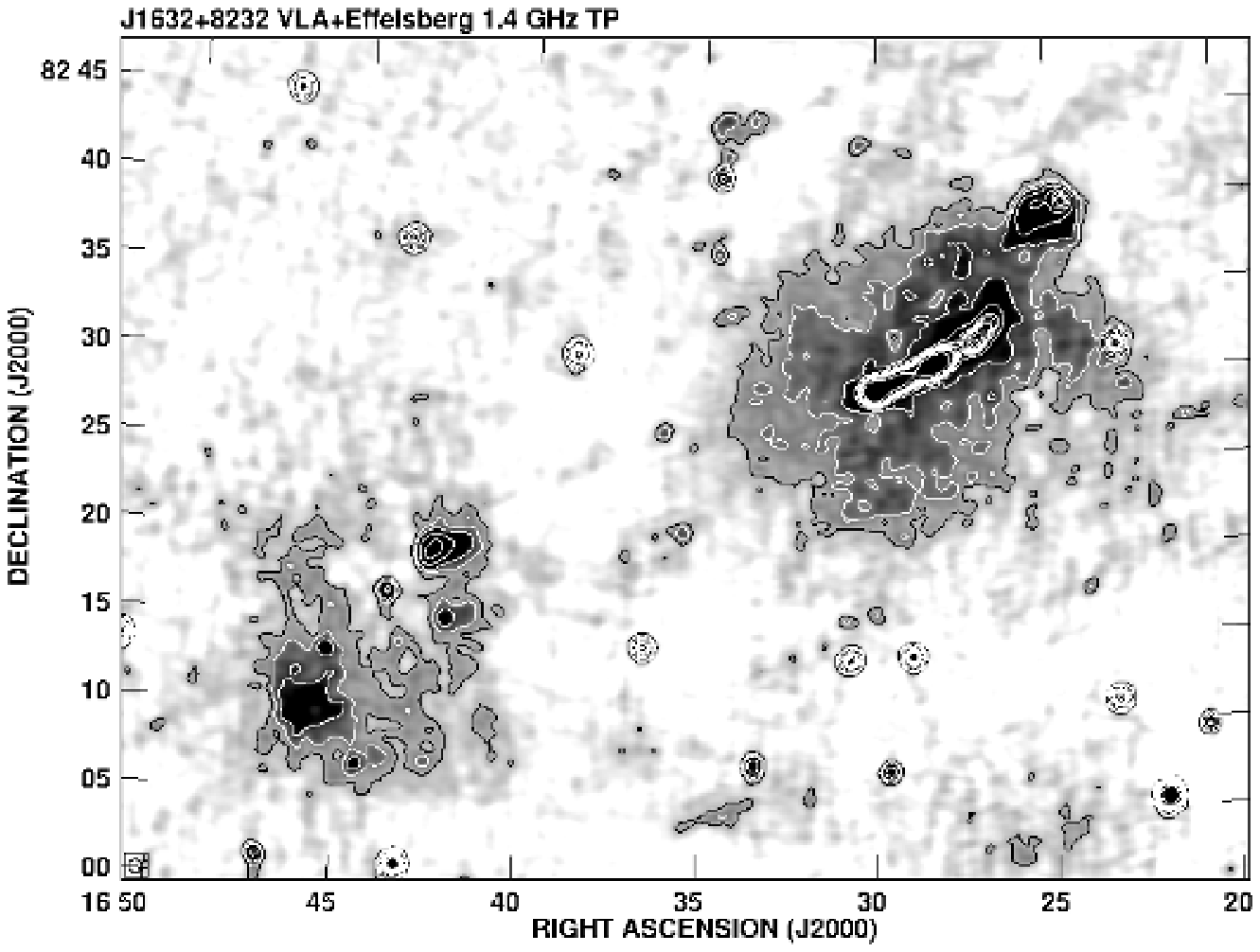}}
\FigCap{GRG J1632+8232. Noise level in both maps is at 0.5~mJy/b.a..}
\end{center}
\label{J1632}
\end{figure}

\subsection{\it GRG J2145+8154}
\label{grg2145}

This source shows the smallest deficit of the radio emission in the NVSS map (Fig.~15, top). As much as 95\% of the merged
map (Fig.~15, bottom) is included. Consequently, both radio maps do not show any diffuse radio emission.

\begin{figure}
\begin{center}
\resizebox{0.7\hsize}{!}{\includegraphics{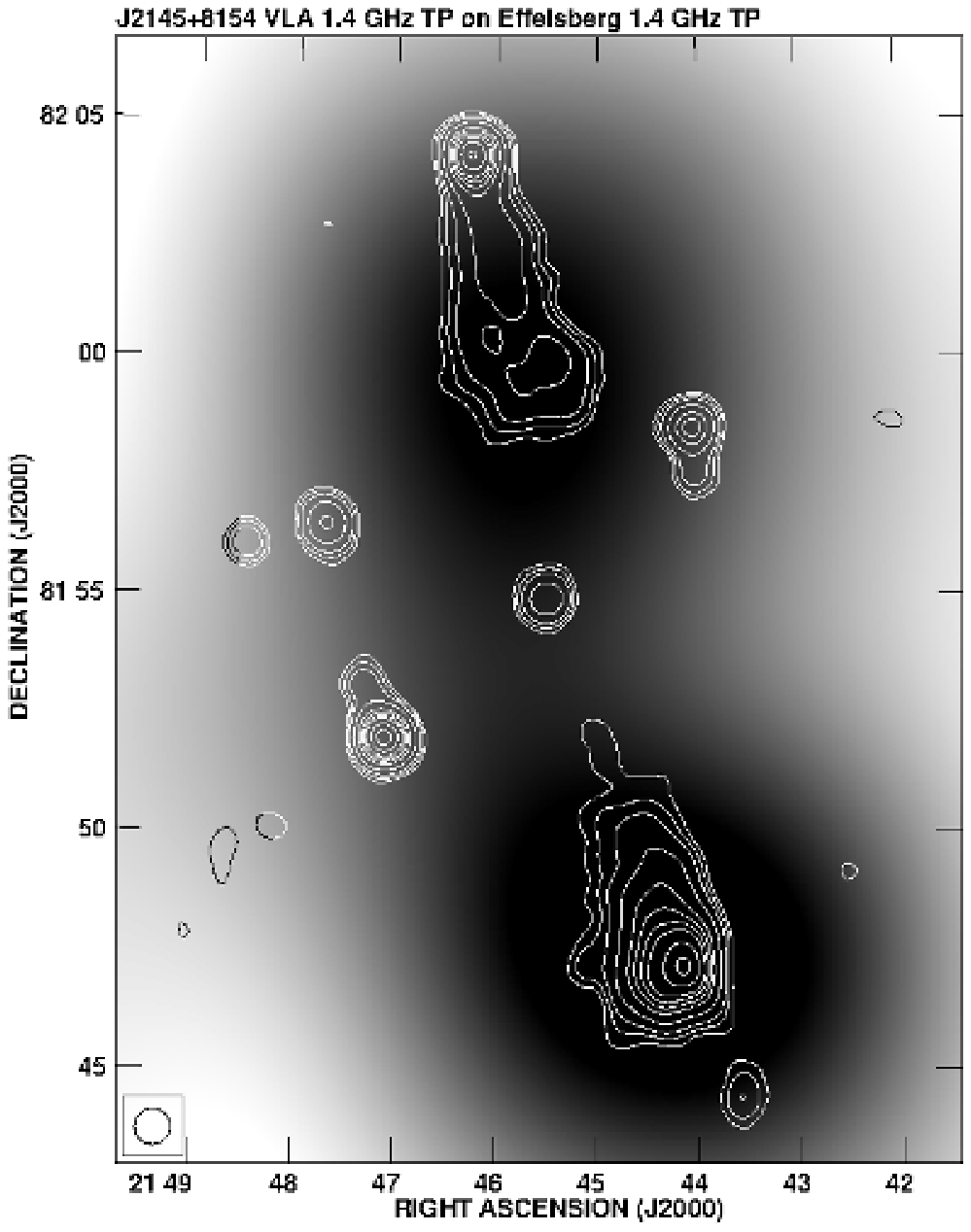}}
\resizebox{0.7\hsize}{!}{\includegraphics{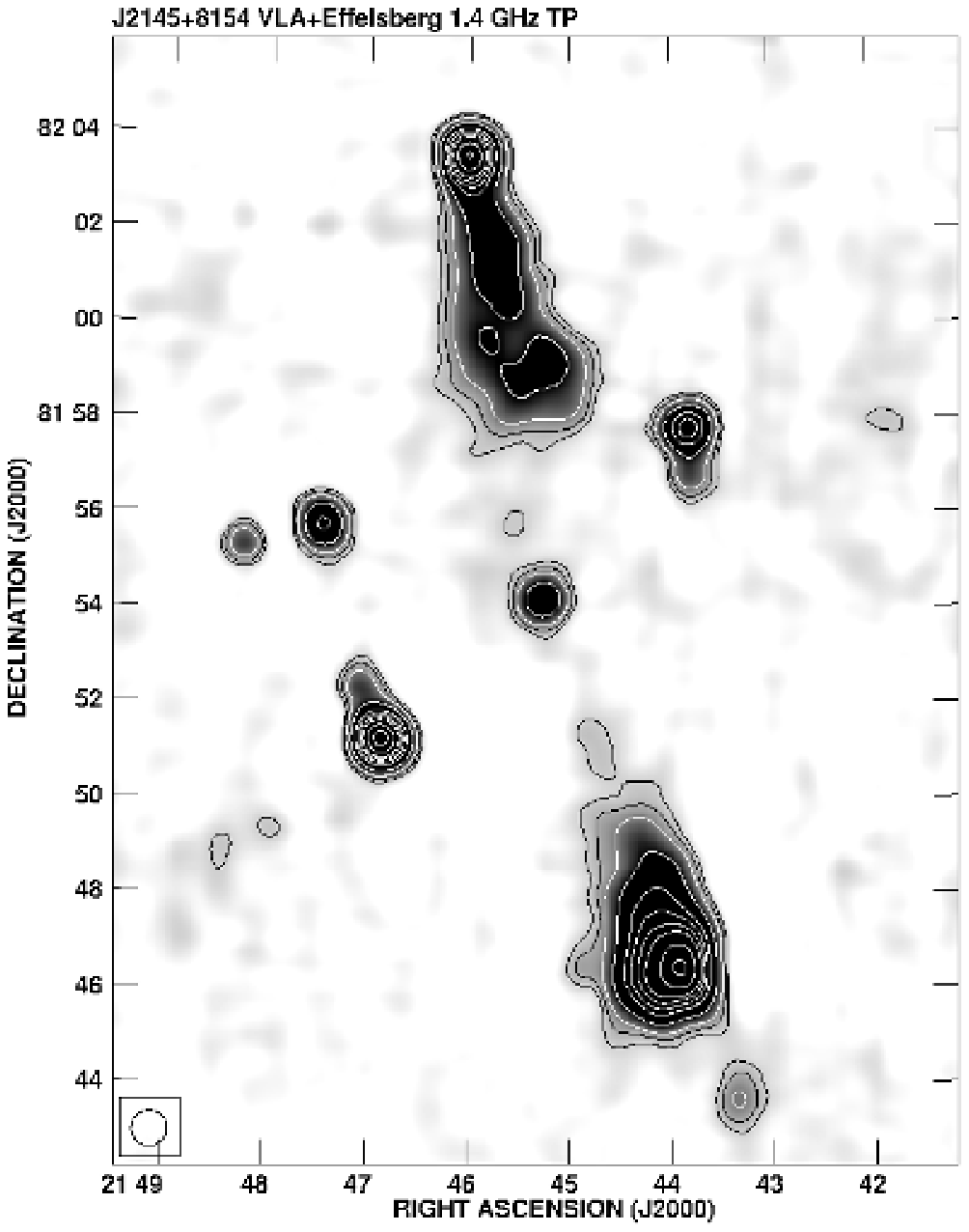}}
\FigCap{GRG J2145+8154. Noise level in both maps is at 0.4~mJy/b.a..}
\end{center}
\label{J2145}
\end{figure}

\FloatBarrier

\begin{landscape}
\begin{table}
\caption{\label{ratios}Comparison of the fluxes between the NVSS and the merged maps}
\begin{center}
\begin{tabular}{llccclc}
\hline\hline
Source          & Other name    &NVSS flux [Jy] & Total flux [Jy]& S$_{NVSS}$/S$_{tot}$ & Halo  & Scale$^{\rm a}$ \\
\hline
J0057+3021      & NGC\,315      & 2.44$\pm$0.05          & 5.05$\pm$0.25           & 0.48                 & No  & 3.63\\
J0107+3224      & 3C\,31        & 4.89$\pm$0.10          & 6.82$\pm$0.34           & 0.72                 & Yes$^{\rm b}$ & 2.50\\
J0318+6829      & WNB\,0313+683 & 0.83$\pm$0.02          & 1.19$\pm$0.06           & 0.70                 & No  & 0.93\\
J0448+4502      & 3C\,129       & 5.73$\pm$0.11          & 7.02$\pm$0.21           & 0.82                 & No  & 1.88\\
J0702+4859      &               & 0.40$\pm$0.01          & 0.52$\pm$0.03           & 0.77                 & No  & 1.19\\
J0748+5548      & DA\,240       & 2.25$\pm$0.05          & 5.46$\pm$0.27           & 0.41                 & Yes & 2.13\\
J0949+7314      & 4C\,73.08     & 2.46$\pm$0.05          & 4.37$\pm$0.22           & 0.56                 & Yes & 0.92\\
J1006+3454      & 3C\,236       & 4.87$\pm$0.10          & 5.93$\pm$0.30           & 0.82                 & No  & 2.44\\
J1032+5644      & HB\,13        & 0.58$\pm$0.01          & 0.84$\pm$0.04           & 0.69                 & No  & 1.16\\
J1312+4450      &               & 0.30$\pm$0.01          & 0.37$\pm$0.02           & 0.82                 & No  & 1.41\\
J1428+2918      &               & 0.44$\pm$0.01          & 0.53$\pm$0.03           & 0.82                 & No  & 0.92\\
J1552+2005      & 3C\,326       & 2.30$\pm$0.05          & 4.24$\pm$0.21           & 0.54                 & Yes & 1.23\\
J1628+5146      & MRK\,1498     & 0.68$\pm$0.01          & 0.76$\pm$0.04           & 0.89                 & No  & 1.15\\
J1632+8232      & NGC\,6251     & 2.18$\pm$0.05          & 4.28$\pm$0.21           & 0.51                 & Yes & 3.25\\
J2145+8154      &               & 0.39$\pm$0.01          & 0.41$\pm$0.02           & 0.95                 & No  & 1.14\\
\hline
\end{tabular}
\end{center}
$^{\rm a}$Angular scale factor - the ratio of the angular size of the source to the theoretical largest scale of the interferometer observations.\\
$^{\rm b}$Probably a radio halo around the entire cluster.\\
\end{table}
\end{landscape}

\section{Discussion}
\label{disc}

\subsection{Global morphologies}
\label{morph}

Although for all galaxies we used the single-dish data that provide the highest sensitivity to diffuse extended radio structures, some of the objects do not show
significant morphology differences when compared to the interferometric NVSS maps. This allows to draw two important conclusions. First of all, it shows that
the method used by us works well in the case of extended diffuse sources. Even though all sources are comparable in size to the single-dish beam, only for those that
show diffuse emission it can be seen in the final maps. In other words, it is a direct proof that no artifact emission is introduced by the large-beam observations
of the single-dish telescope when merging with the high-resolution data. Some of the sources showed large-scale diffuse radio structures that might be directly
associated with past periods of radio activity. Radio galaxies with signatures of previous activity episodes are still being reported~(e.g. Saikia and Jamrozy~2009).
Out of 15 objects in our sample, 5 GRGs show a distinct radio halo/cocoon structure. Three of them are observed around
sources whose FR type cannot be clearly determined (see Tables~1 and~2). One is found around an FRII source and another around an
FRI source, provided that the halo around 3C\,31 is a cocoon and not a radio halo of the cluster (see Sect.~3.2 for details). Therefore,
in our study we are not in position to draw clear conclusions about any relations between the existence of a radio cocoon around a source and its FR type. Observations
of a larger sample of objects (which would, unfortunately, also include sources with a smaller angular extent) are certainly needed to establish/exclude
such relations.

By using the single-dish data in our study we were able to observe the entire extent of any possible diffuse radio emission in the maps of our sample.
This means that any information acquired is limited only by the observing frequency of 1.4\,GHz and physical properties of the sources.
The distributions of diffuse emissions resemble those observed at much lower frequencies.
A direct proof comes from such comparison of the radio emission from DA\,240 giant radio galaxy. Both 1.4\,GHz (lower panel of Fig.~6)
and 609\,MHz~(Fig.~13 in Mack et al.~1997) radio maps show almost exactly the same extent of the radio halo. The most important issue, however, is
a very similar morphology of the cocoon in both maps, proving that the large-scale emission in the 1.4\,GHz merged map cannot be an artificial signal
introduced by the merging process and is just the actual emission from the source.
As usually the extent of the low surface-brightness emission from GRGs increases with wavelength,
this clearly shows the need for single-dish observations when extended diffuse radio emission from angularly large sources
is concerned. It also points at poor sensitivity to large-scale structures of the low-frequency interferometric observations. This is most likely the main
reason for the absence of any much extended emission in the low-frequency interferometer maps, which would be expected when taking into account steep spectral
indices. We note that combining single-dish and interferometric maps at lower frequencies would help to trace such emission. The main problem, however,
arises from very large beams of single-dish radio telescopes at frequencies of hundreds of MHz, which would make combination of the data very difficult, with
an increasing number of unrelated sources found within the observing beam.

Table~2 summarizes losses of flux due to interferometric observations. For each of the sample sources, the fluxes from both the merged map
and the NVSS map are provided and their ratios calculated. The last two columns of the table indicate the presence or not of the halo around the source and compare its angular extent to the largest angular scale available to the interferometer observations.

For the VLA radio telescope, the most compact D-configuration allows to observe at 1.4\,GHz angular coherent scales not larger than 16$\arcmin$.
This is valid, however, only for a full synthesis observations of 8 hours. For much shorter integrations, as in the case of the NVSS data, its size can be limited
as much as to half of this value. Still, even taking into account this higher theoretical angular scale of 16$\arcmin$ we can see that this exceeds practically all of the sources in our sample. Thus it would be natural to expect that for all sources the flux loss would increase with their size.
Table~2 shows that although for most of the sources the flux loss is significant, it cannot be simply related to the size of the source.
We can, however, observe a more tight correlation of the flux loss with a presence of radio halo/cocoon around a source.
If we compare both dependencies, we can notice that there are objects in our sample which show both a smaller
angular size, which allows them, at least theoretically, to match the scale observed by the VLA radio telescope, and a significant flux loss due to the presence of
radio cocoon. It is most clearly visible for 4C\,73.08 (Fig.~7) and 3C\,326 (Fig.~12) giant radio galaxies.
As mentioned before, both sources can be suspected of previous jet activity that results in significant diffuse radio emission forming an extended
cocoon around the more bright and recent emission from the central source and its well-defined radio lobes.
Subrahmanyan et al.~(1996) suggested that GRGs may have attained their large sizes as a result of restarting jet activity.
Since, as it was mentioned above, the scale of observed
extended emission depends also on sensitivity, we can infer that low surface-brightness diffuse radio cocoons around giant radio galaxies are practically
invisible to radio interferometers. This supports also the fact mentioned in Sect.~3 where each source is discussed, that the brightnesses and morphologies
of radio halos to be seen in the merged Effelsberg+VLA maps at 1.4\,GHz are similar to those visible in the maps at 325\,MHz from the WSRT radio telescope. Given a typical
spectral index of radio galaxy diffuse emission, we should expect much brighter and more extended emission at lower frequencies. These, however, are
hardly accessible to single-dish observations because of very large beams of the order of $1\degr$ or more.

Therefore, sensitive high-resolution low-frequency observations with the use of new-generation observatories like
Low-Frequency Array (LOFAR), Long Wavelength Array (LWA), or Murchinson Widefield Array (MWA)
could help to detect many radio halos around other, also more distant, radio galaxies.

\subsection{Spectral analysis}
\label{ageing}

Table~3 presents the 1.4\,GHz fluxes obtained from the merged NVSS+Effelsberg maps along with the measurements at other frequencies taken from the
literature, or, like in the case of some observations at 0.3\,GHz and 5\,GHz, from the available fits data. All measurements were used
to construct a global spectral index for each source. To check consistency of the data, we performed a spectral fit with the {\em SYNAGE} algorithm
of~Murgia~(1996). We used the~Jaffe and Perola~(1973) model, treating $\alpha_{\rm inj}$ and the break frequency as free parameters
(for details of the model fitting, see e.g.~Jamrozy et al.~2008). The results are given in Figs.~16 to~19.
In all plots, along with the flux density value at 1.4\,GHz measured in our merged maps, we marked the NVSS flux density value with an open symbol. This helps to
show the influence of the missing flux on the integrated spectrum. We note here that many of the fluxes at frequencies higher than 1.4\,GHz are single-dish measurements,
while all fluxes at lower frequencies are interferometric measures (see Table~3 and discussion in Sect.~4.1).
We did not attempt to derive the synchrotron ages of these objects, because the flux density for an entire source
originates from areas which differ substantially in terms of physical conditions (i.e. magnetic field strength, etc.).

For 8 sources from our sample we obtained reasonable values for the break frequency $\nu_{\rm break}$ of few tens of GHz. This, of course, calls for sensitive observations
at centimetre wavelengths to verify our findings. Nevertheless, as one can expect, the possibility of determination of the break frequency is closely related
to the presence of diffuse radio structures around the source. From that we may infer, that relatively low break frequencies are specific for radio sources with cocoons,
that very likely are signatures of past cycles of jet activity. The diffuse radio emission shows often steep spectra and therefore contributes significantly
to the ageing of the source. For younger objects, with bright centres and hot spots that have flatter spectra, the break frequency might be difficult to determine
or the result is some unphysical value. For some sources, however, both conditions can be fulfilled, as in the case of DA\,240, which shows an impressive radio cocoon but
its break frequency is determined as unrealistically high. This shows that a detailed ageing analysis for such sources is difficult, as it requires a clear separation
of the radio source's components, i.e. centres, hot spots, halos. Although possible at higher frequencies, this meets severe problems at the low end of radio frequencies,
where either high sensitivity or high resolution of the observations is missing, as we mentioned in Sect.~4.1.

\begin{landscape}
\begin{table}
\caption{\label{fluxes1}Flux densities (in Jy) of the sample giant radio galaxies}
\begin{center}
\begin{tabular}{crrrrrr}
\hline\hline
Source          & 0.3\,GHz              & 0.6\,GHz              & 1.4\,GHz              & 2.7\,GHz              & 5\,GHz                & 10.5\,GHz             \\
\hline
J0057+3021      &$9.71\pm0.12^{\rm 1}$  &$5.33\pm0.10^{\rm 1}$  &$5.05\pm0.25^{\rm 3}$&$3.37\pm0.09^{\rm 1}$    &$2.46\pm0.07^{\rm 1}$  &$2.19\pm0.08^{\rm 1}$  \\
                &                       &$6.29\pm0.20^{\rm 2}$  &                       &                       &                       &                       \\
J0107+3224      &$13.67\pm0.28^{\rm 4}$ &$9.51\pm0.14^{\rm 2}$  &$6.82\pm0.34^{\rm 3}$&$3.53\pm0.25^{\rm 5}$    &$1.94\pm0.10^{\rm 6}$  &$1.02\pm0.09^{\rm 5}$  \\
                &                       &                       &                       &                       &$2.09\pm0.07^{\rm 5}$  &                       \\
J0318+6829      &$3.14\pm0.15^{\rm 7}$  &                       &$1.19\pm0.06^{\rm 3}$  &                       &$0.34\pm0.04^{\rm 7}$  &$0.23\pm0.01^{\rm 7}$  \\
J0448+4502      &$21.20\pm1.40^{\rm 8}$ &$14.50\pm1.40^{\rm 8}$ &$7.02\pm0.21^{\rm 3}$&                   &$2.38\pm0.12^{\rm 6}$  &                       \\
J0702+4859      &$1.18\pm0.09^{\rm 4}$  &                       &$0.52\pm0.03^{\rm 3}$  &                       &$0.18\pm0.02^{\rm 6}$  &                       \\
J0748+5548      &$17.05\pm0.20^{\rm 1}$ &$9.23\pm0.13^{\rm 1}$  &$5.46\pm0.27^{\rm 3}$  &$2.71\pm0.06^{\rm 1}$  &$1.77\pm0.04^{\rm 1}$  &$1.04\pm0.05^{\rm 1}$  \\
J0949+7314      &$10.43\pm0.21^{\rm 9}$ &$7.28\pm0.13^{\rm 2}$  &$4.37\pm0.22^{\rm 3}$  &                       &$1.64\pm0.15^{\rm 9}$  &$0.61\pm0.02^{\rm 9}$  \\
J1006+3454      &$13.13\pm0.14^{\rm 1}$ &$8.23\pm0.91^{\rm 1}$  &$5.93\pm0.30^{\rm 3}$  &$3.65\pm0.07^{\rm 1}$  &$2.35\pm0.04^{\rm 1}$  &$1.27\pm0.03^{\rm 1}$  \\
                &                       &$9.17\pm0.23^{\rm 2}$  &                       &                       &                       &                       \\
J1032+5644      &$1.09\pm0.06^{\rm 4}$  &$1.09\pm0.10^{\rm 2}$  &$0.84\pm0.04^{\rm 3}$  &                       &$0.27\pm0.02^{\rm 6}$  &                       \\
J1312+4450      &$0.65\pm0.03^{\rm 4}$  &                       &$0.37\pm0.02^{\rm 3}$  &                       &$0.16\pm0.02^{\rm 6}$  &                       \\
J1428+2918      &$1.15\pm0.04^{\rm 9}$  &                       &$0.53\pm0.03^{\rm 3}$  &                       &$0.19\pm0.04^{\rm 9}$  &$0.08\pm0.01^{\rm 9}$  \\
J1552+2005      &$11.96\pm0.13^{\rm 1}$ &$7.40\pm0.11^{\rm 1}$  &$4.24\pm0.21^{\rm 3}$  &$2.22\pm0.07^{\rm 1}$  &$1.30\pm0.01^{\rm 1}$  &$0.43\pm0.02^{\rm 1}$  \\
J1628+5146      &$1.62\pm0.03^{\rm 9}$  &                       &$0.76\pm0.04^{\rm 3}$  &                       &$0.30\pm0.03^{\rm 9}$  &$0.25\pm0.01^{\rm 9}$  \\
J1632+8232      &$11.55\pm0.18^{\rm 1}$ &$6.93\pm0.12^{\rm 1}$  &$4.28\pm0.21^{\rm 3}$  &                       &                       &$1.55\pm0.06^{\rm 1}$  \\
J2145+8154      &$1.06\pm0.05^{\rm 9}$  &                       &$0.41\pm0.02^{\rm 3}$  &                       &                       &$0.11\pm0.01^{\rm 9}$  \\
\hline
\end{tabular}
\end{center}
References: $^{\rm 1}$Mack et al.~1997; $^{\rm 2}$J\"{a}gers~PHD; $^{\rm 3}$This work; $^{\rm 4}$WENSS - Rengelink et al.~1997 (original data measured by us);
$^{\rm 5}$Laing \& Peacock~1980; $^{\rm 6}$87GB survey - Condon et al.~1989 (original data measured by us); $^{\rm 7}$Schoenmakers et al.~1998; $^{\rm 8}$Lal \& Rao~2004;
$^{\rm 9}$Schoenmakers et al.~2000.\\
\end{table}
\end{landscape}

\begin{figure}
\begin{center}
\resizebox{0.45\hsize}{!}{\includegraphics{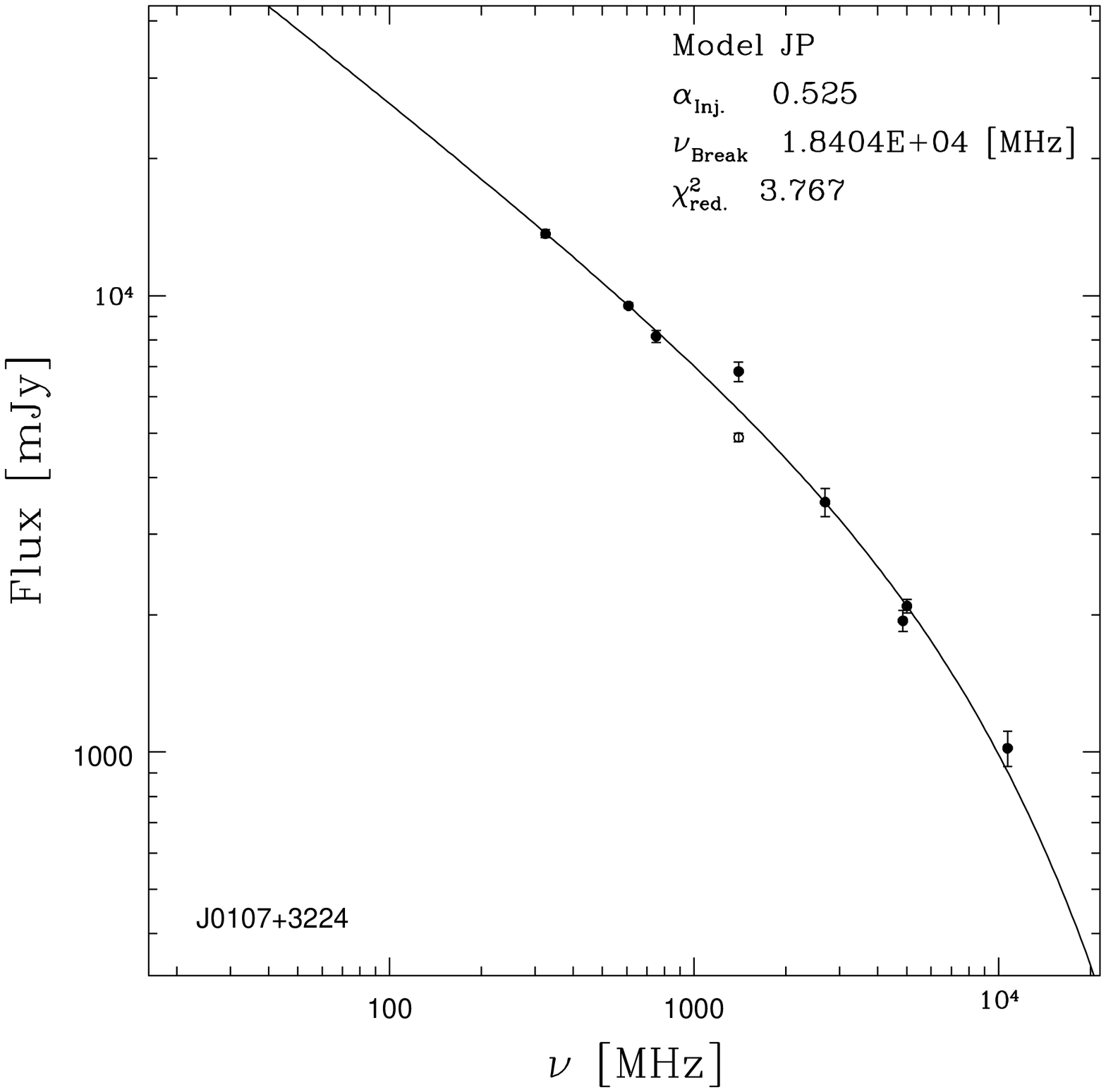}}
\resizebox{0.45\hsize}{!}{\includegraphics{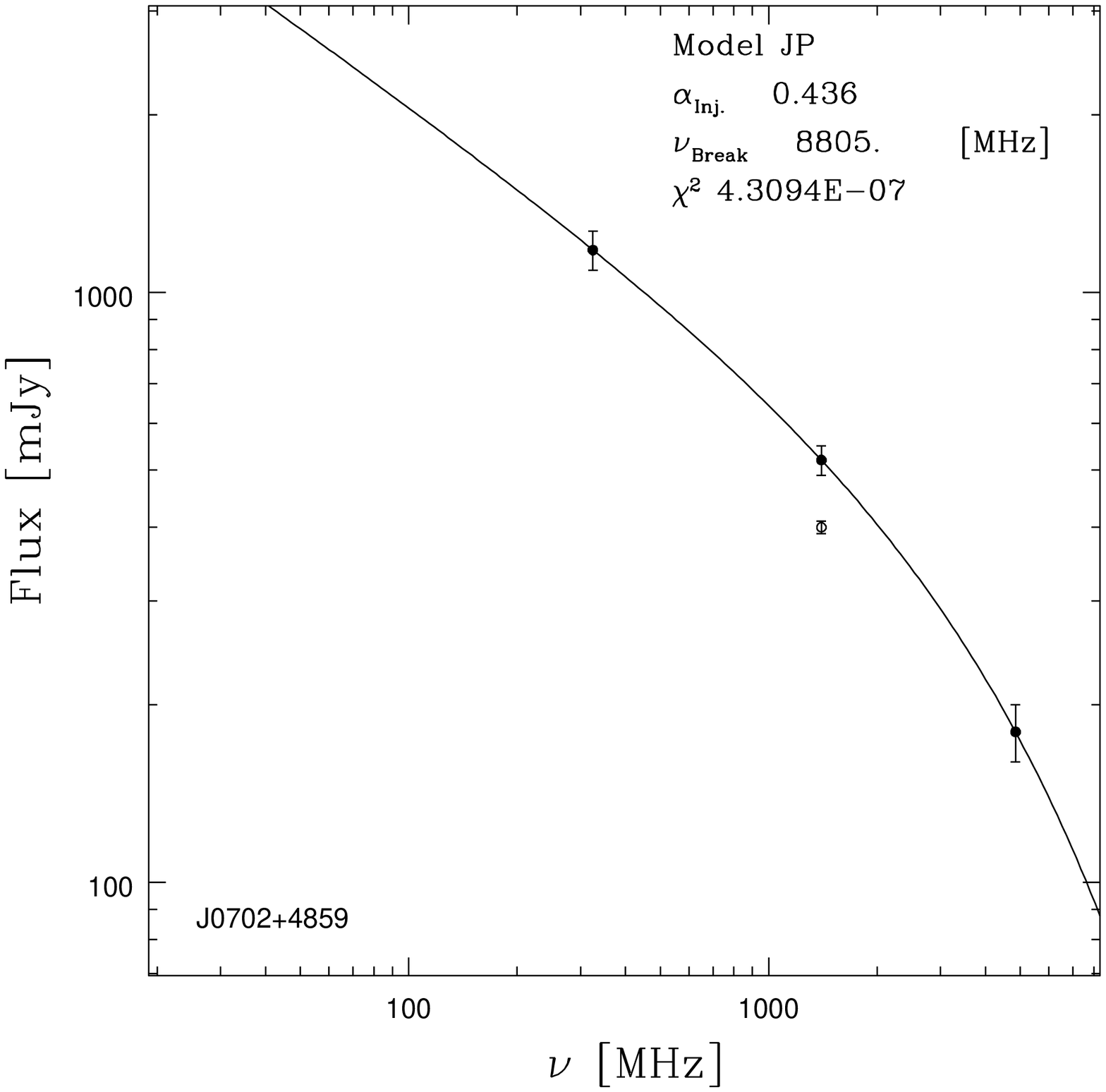}}
\resizebox{0.45\hsize}{!}{\includegraphics{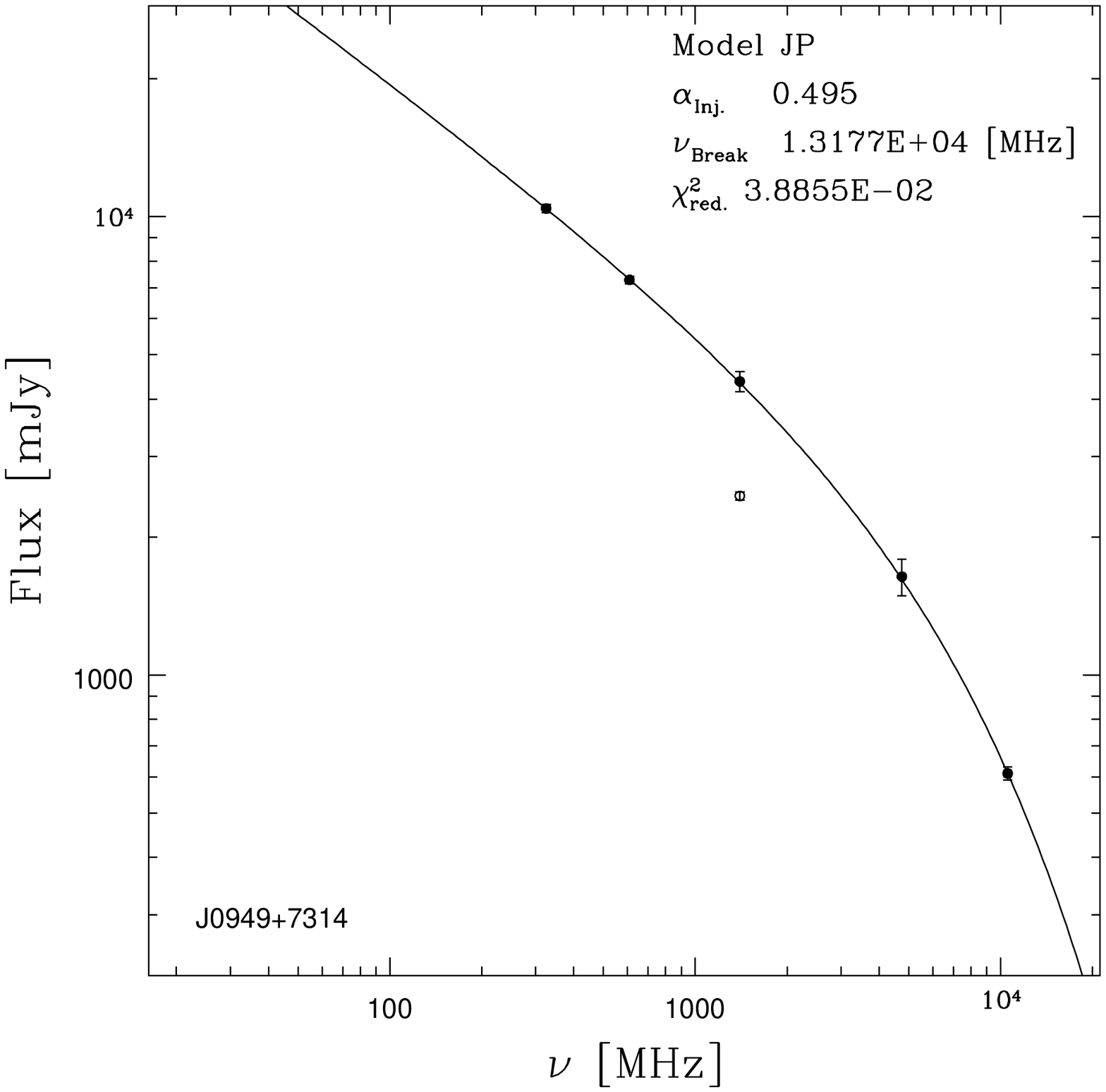}}
\resizebox{0.45\hsize}{!}{\includegraphics{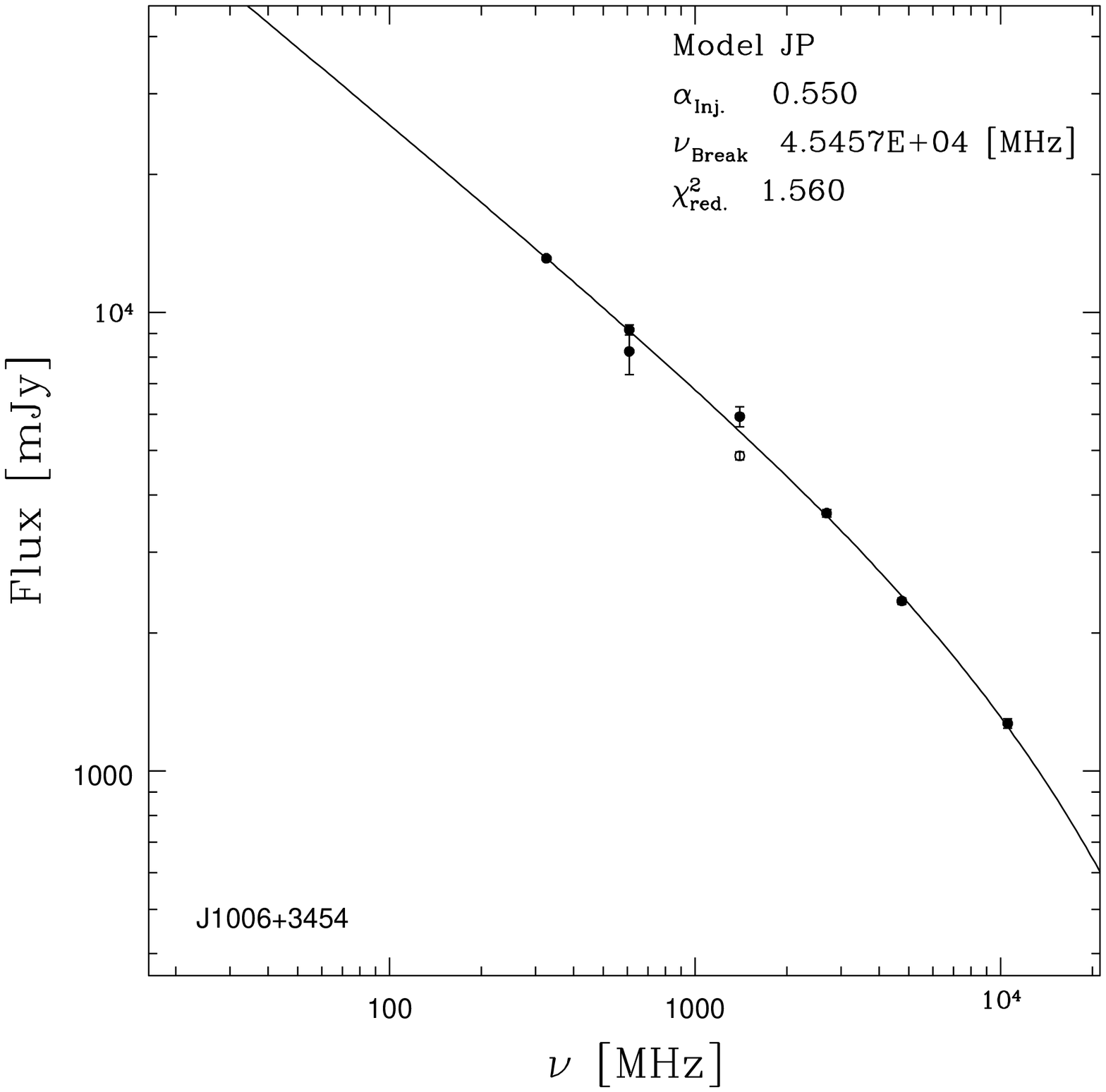}}
\resizebox{0.45\hsize}{!}{\includegraphics{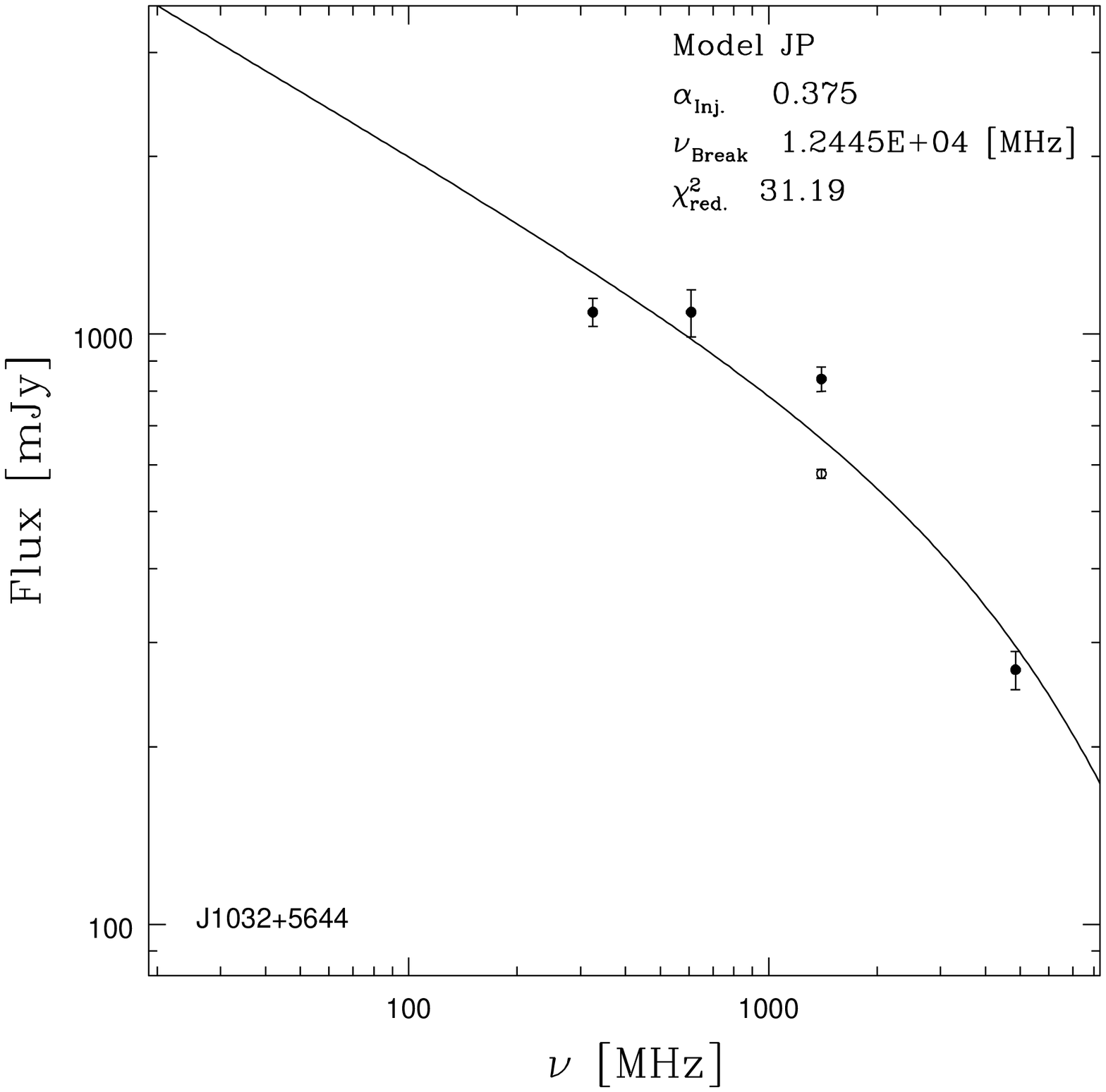}}
\resizebox{0.45\hsize}{!}{\includegraphics{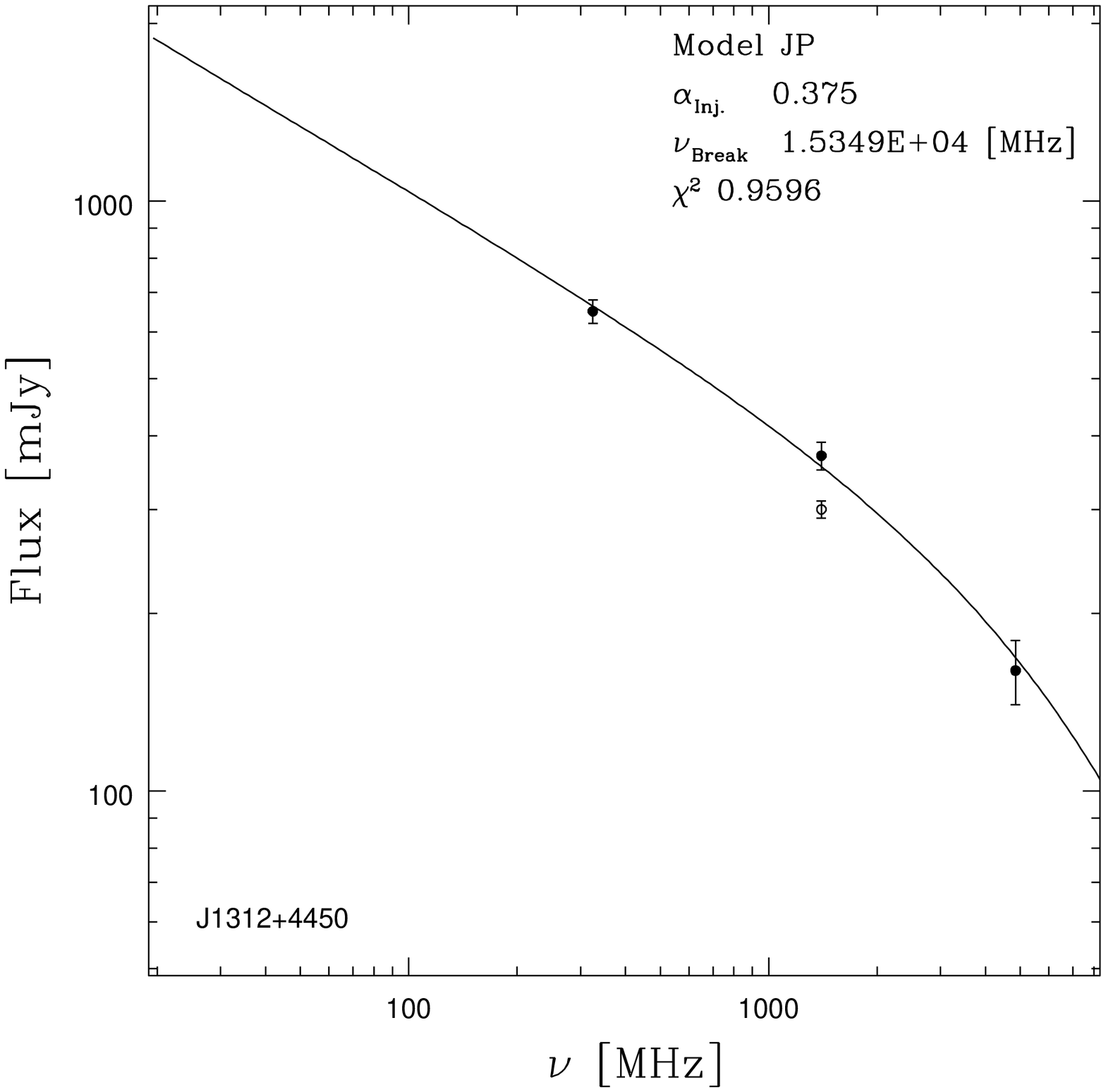}}
\end{center}
\FigCap{Global spectra of sources, which have curved spectra and the breaking frequency $\nu_{\rm break}$ is located in the GHz radio regime.}
\label{break}
\end{figure}

\begin{figure}
\begin{center}
\resizebox{0.45\hsize}{!}{\includegraphics{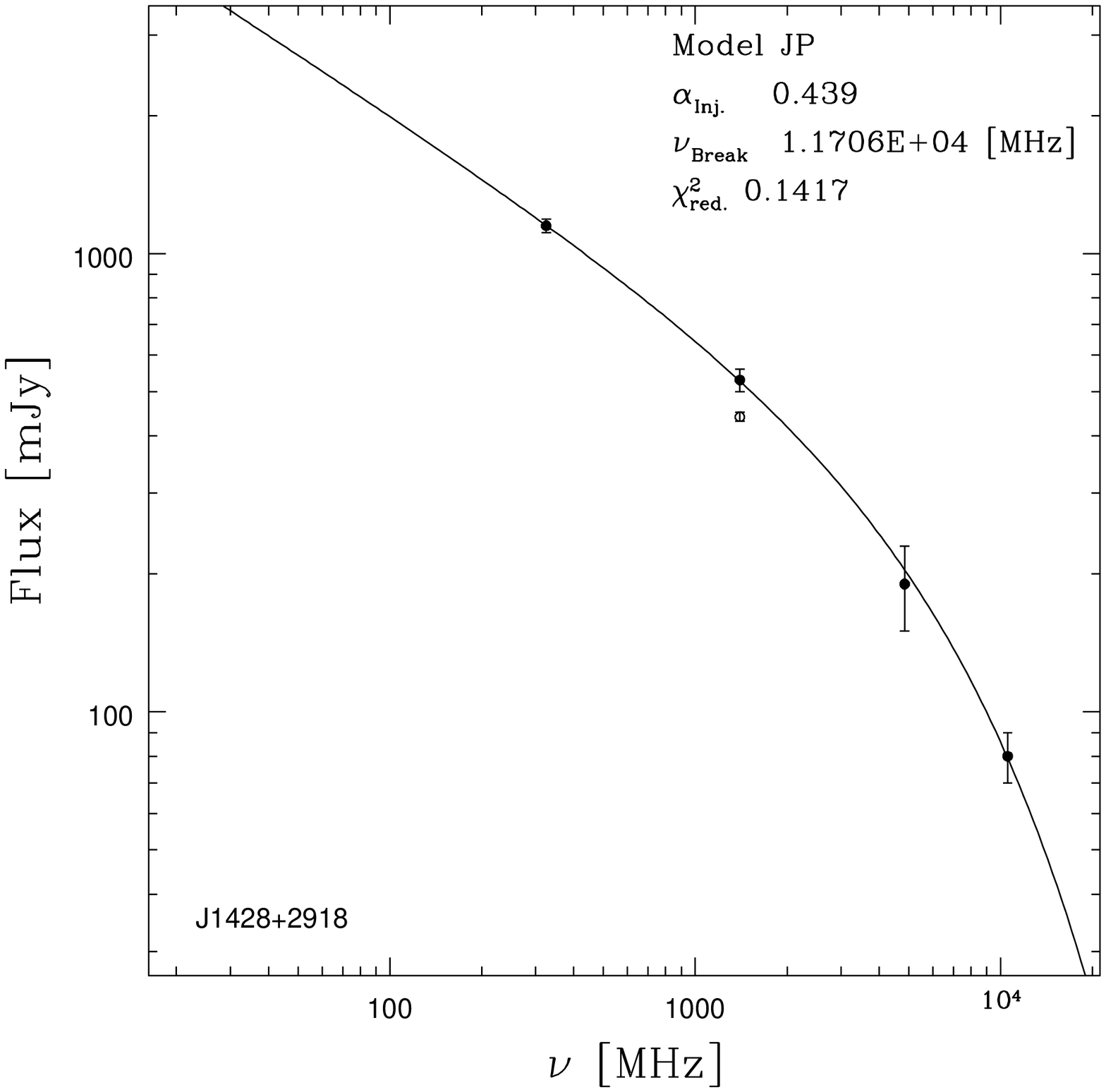}}
\resizebox{0.45\hsize}{!}{\includegraphics{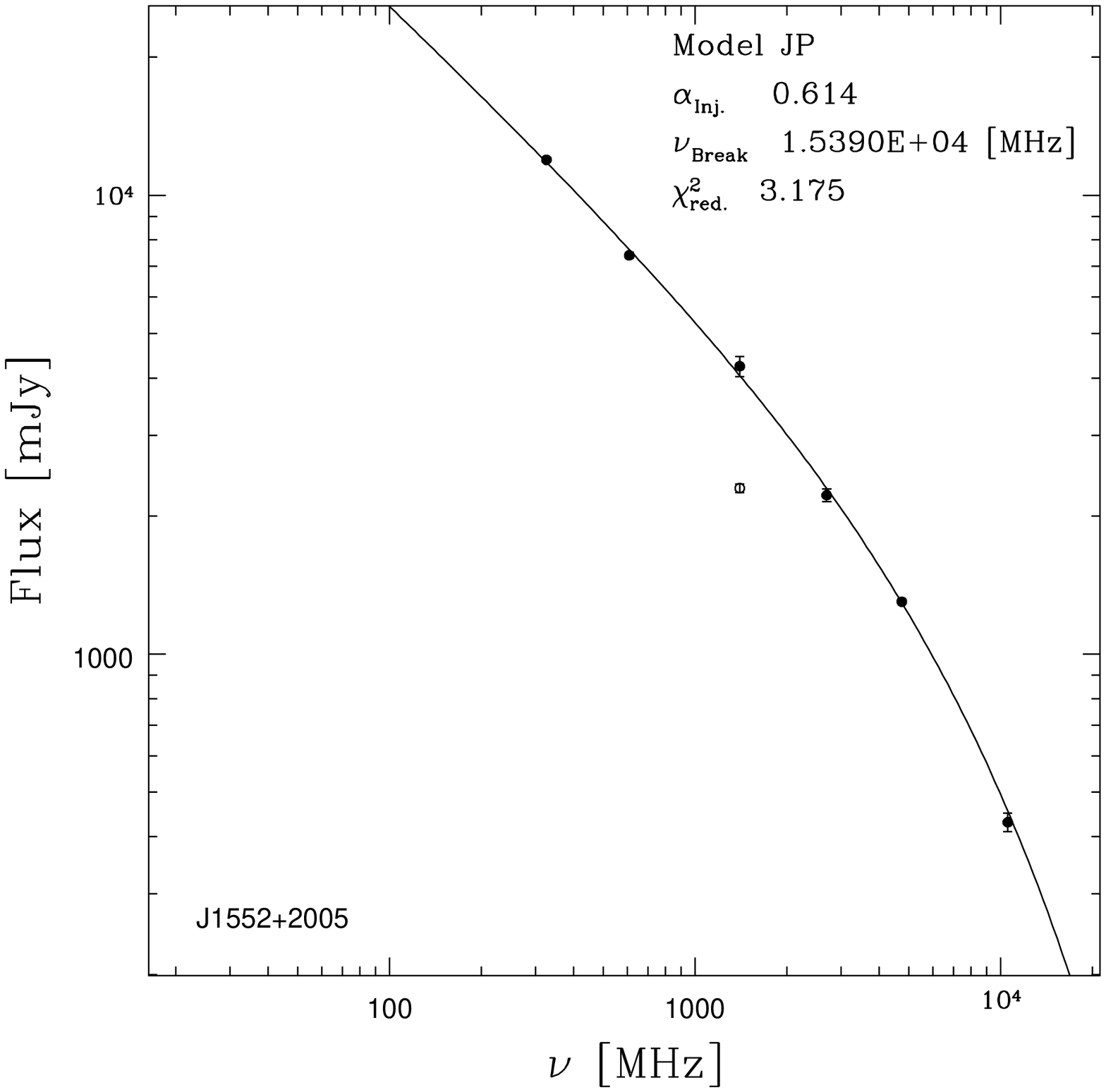}}
\end{center}
\FigCap{Continuation of Fig.~16.}
\label{break2}
\end{figure}

\begin{figure}
\begin{center}
\resizebox{0.45\hsize}{!}{\includegraphics{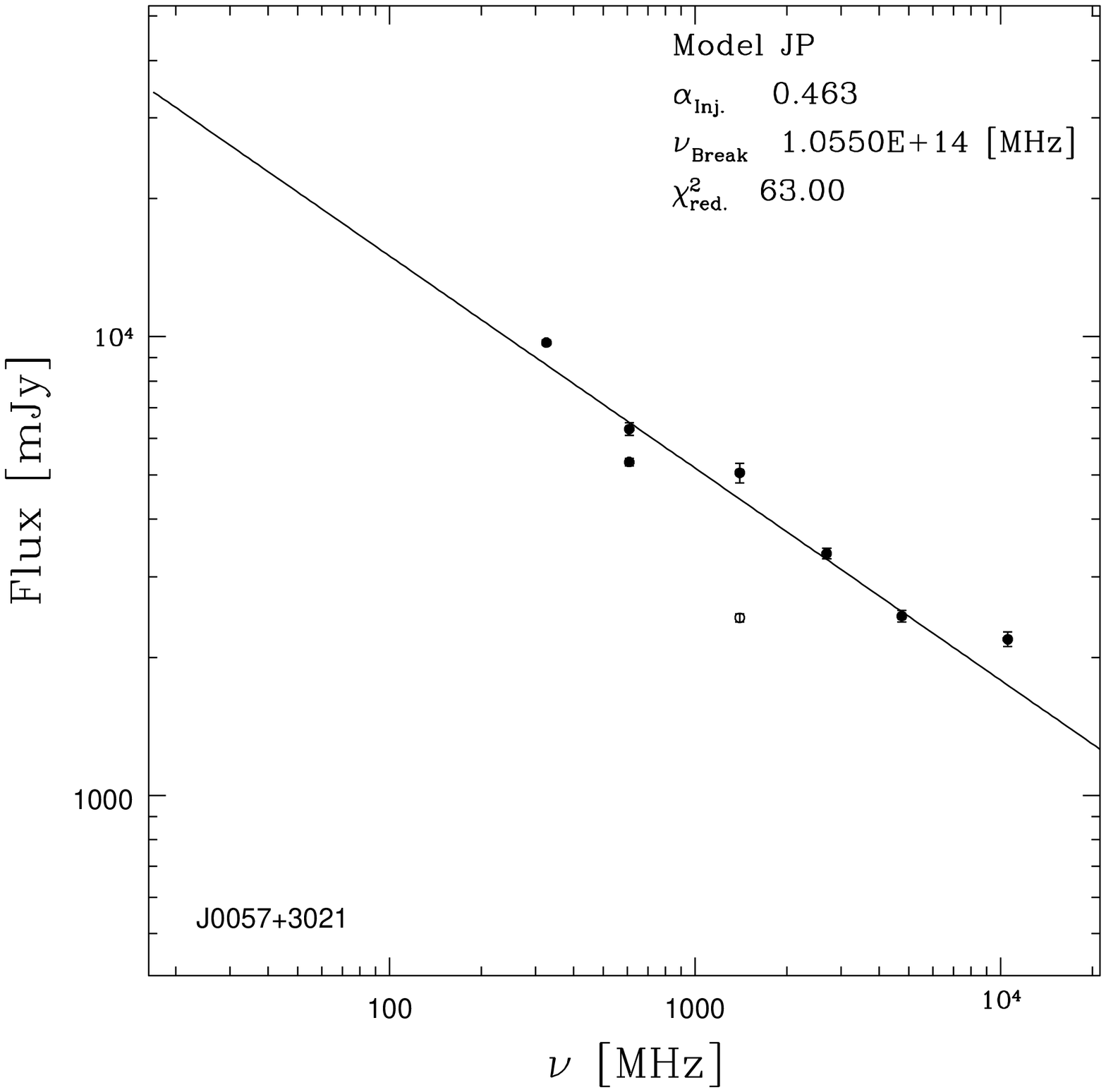}}
\resizebox{0.45\hsize}{!}{\includegraphics{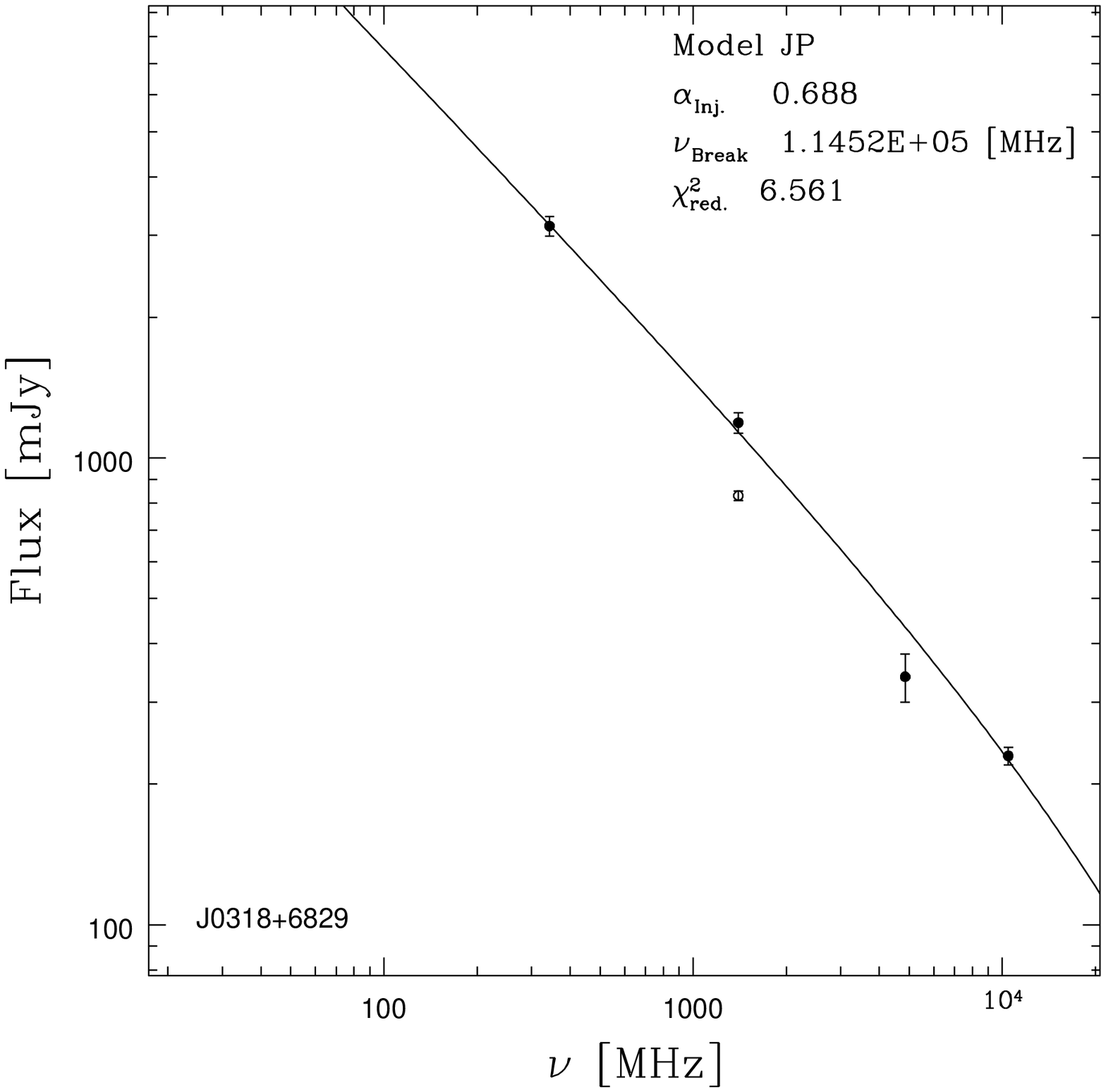}}
\resizebox{0.45\hsize}{!}{\includegraphics{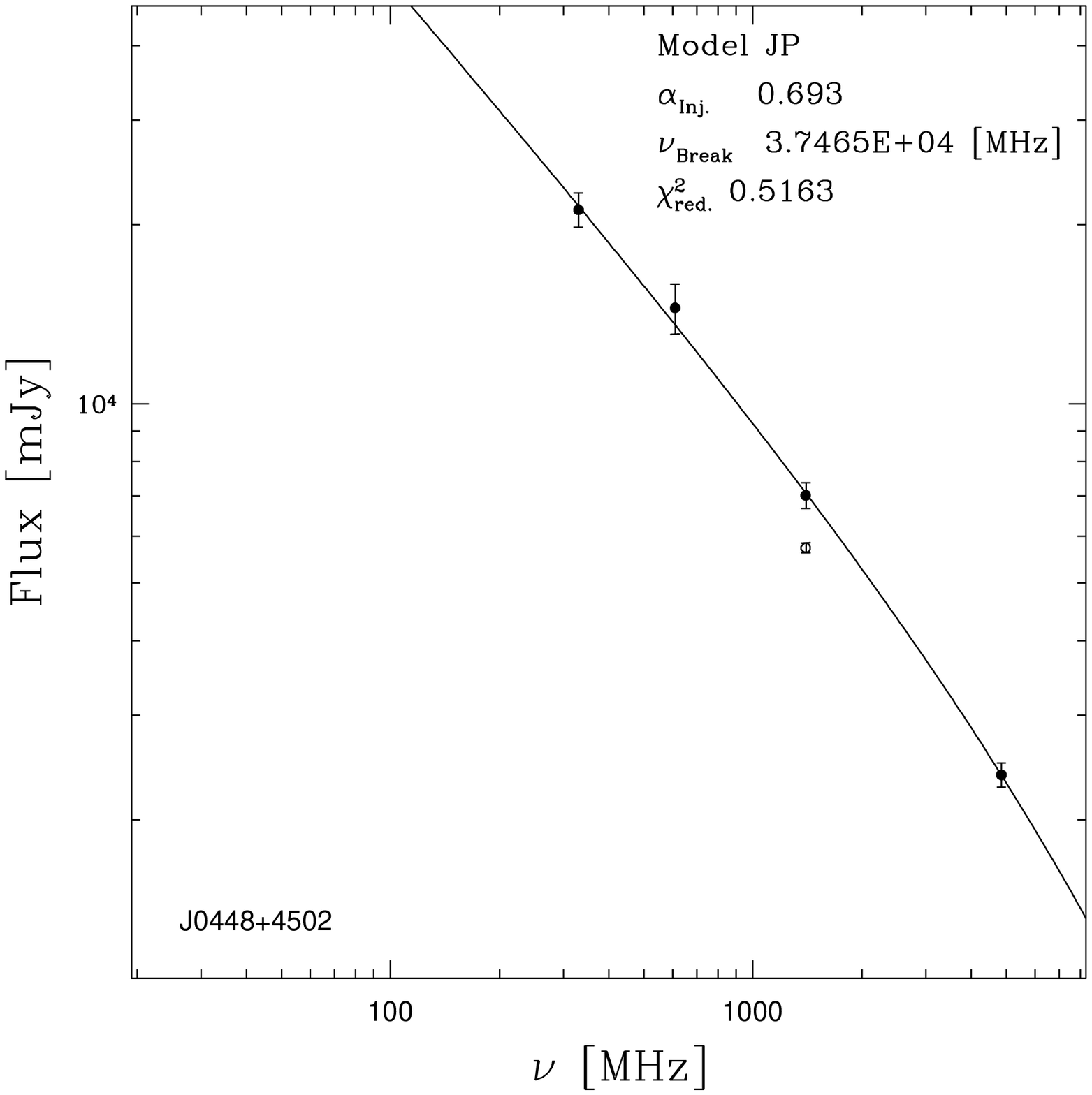}}
\resizebox{0.45\hsize}{!}{\includegraphics{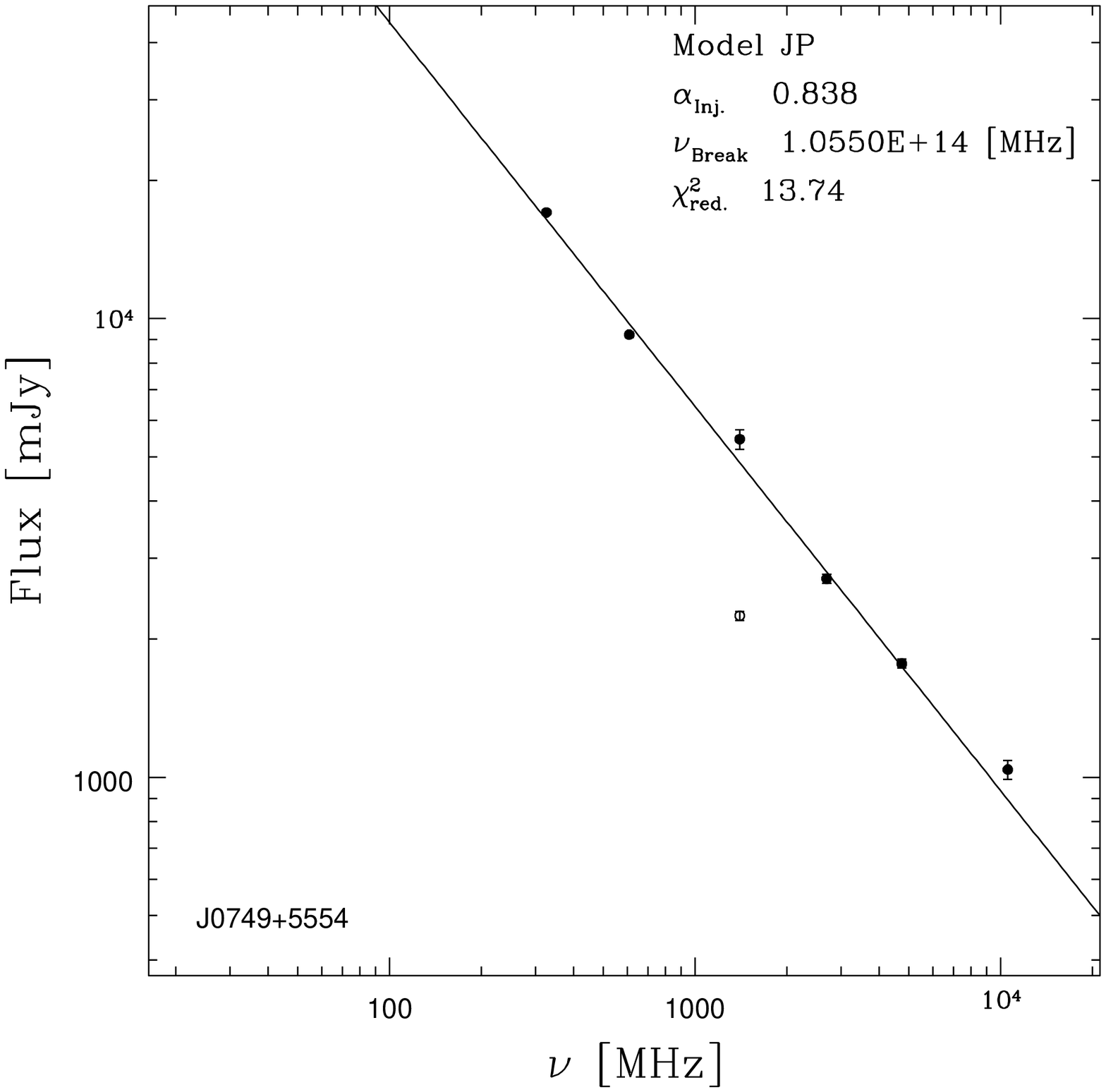}}
\end{center}
\FigCap{Global spectra of sources, which do not show any curvature in the GHz radio regime.}
\label{nobreak}
\end{figure}

\begin{figure}
\begin{center}
\resizebox{0.45\hsize}{!}{\includegraphics{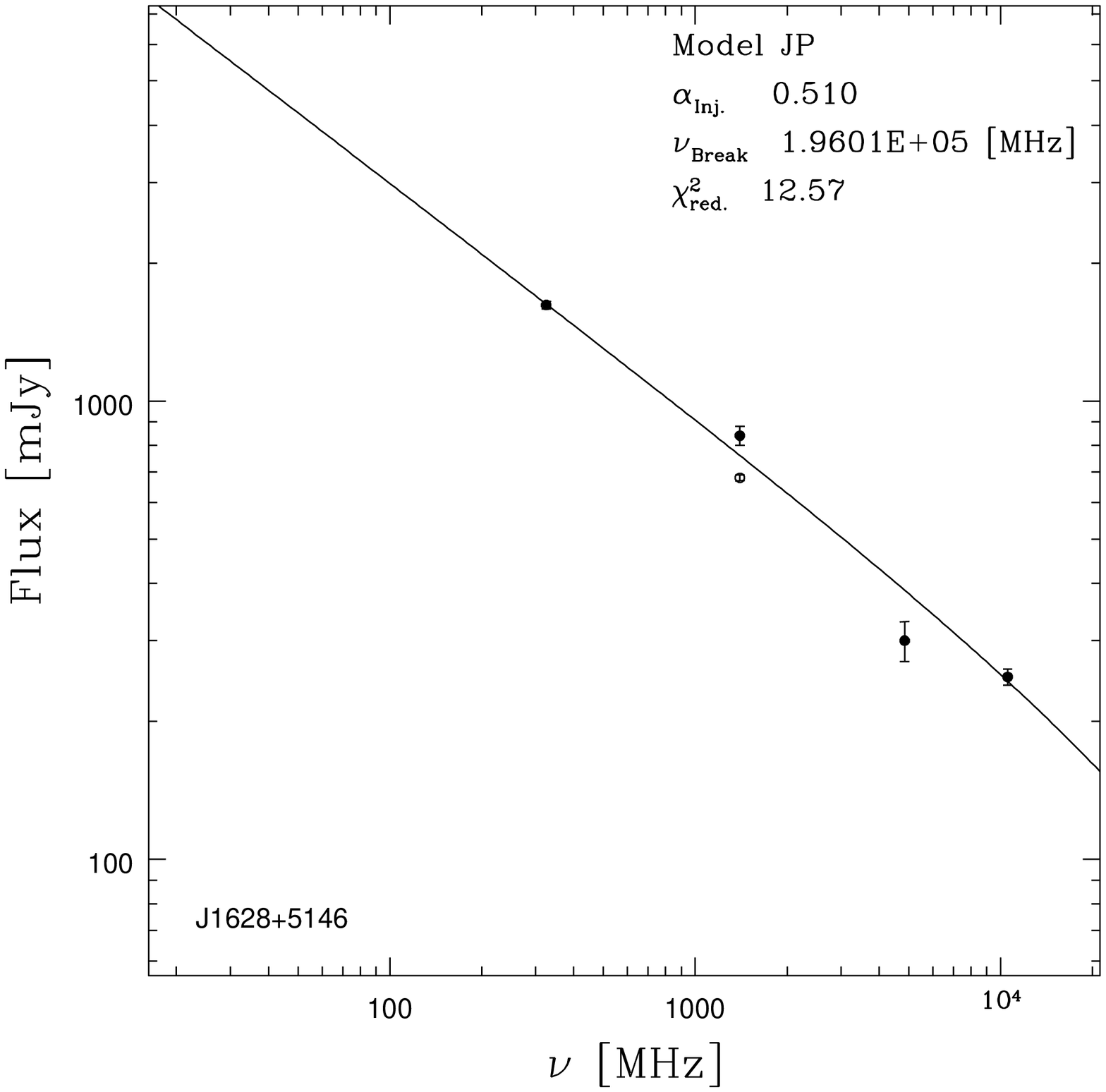}}
\resizebox{0.45\hsize}{!}{\includegraphics{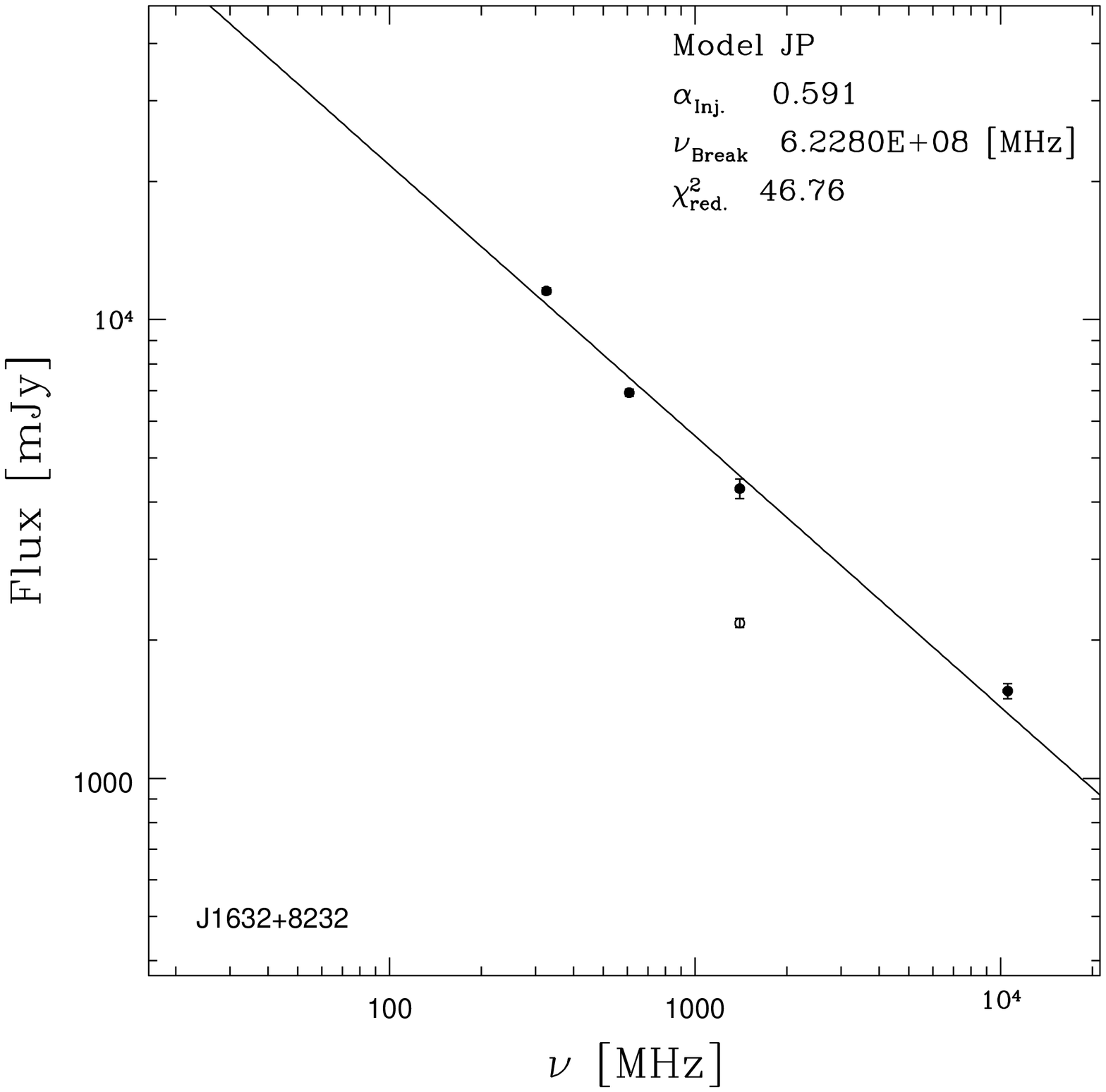}}
\resizebox{0.45\hsize}{!}{\includegraphics{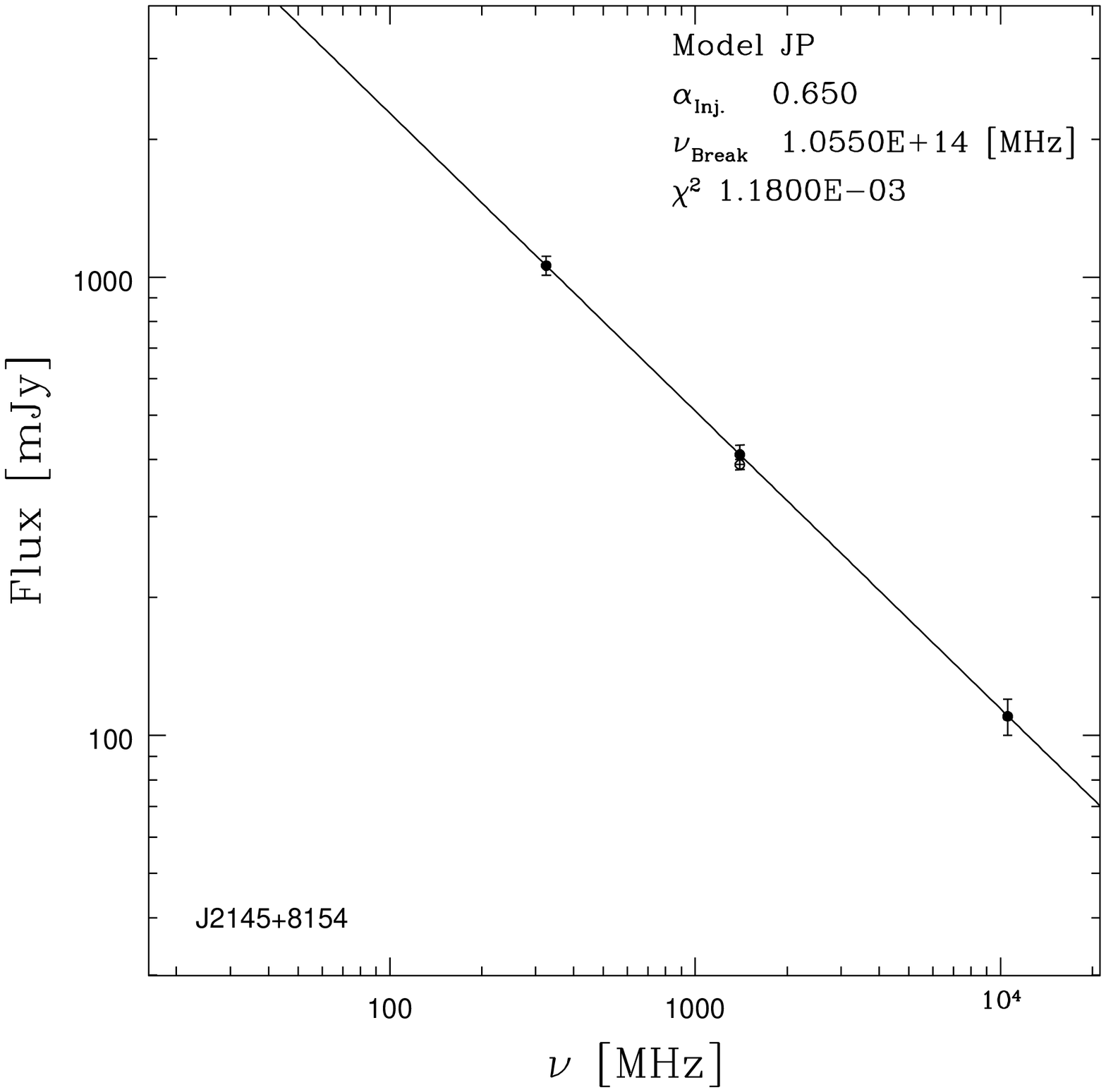}}
\end{center}
\FigCap{Continuation of Fig.~18.}
\label{nobreak2}
\end{figure}

\FloatBarrier

\section{Conclusions}

The detailed study of the radio maps of a sample of 15 giant radio galaxies allowed us to conclude the following.

\begin{itemize}
\item[-] For the extended radio sources short NVSS observations provide data that often show a significant underestimation of the total flux.
\item[-] Merging interferometric and single-dish radio data for extended sources yields high-resolution sensitive maps.
These maps sometimes reveal low surface-brightness diffuse emission, which suggests a significant deficit of large-scale emission
at low frequencies, where only interferometric observations can provide high-resolution data for the giant radio galaxies.
\item[-] Single-dish observations of extended radio galaxies are crucial for a proper flux determination, as well as to determine
a possible presence of a radio cocoon around the source.
\item[-] Ageing analysis is possible mainly for giant radio galaxies with cocoons. It is, however, difficult to estimate contribution
of diffuse emission to the total flux, due to resolution and/or sensitivity problems at low frequencies.
\item[-] 5 of the 15 giant radio galaxies reveal extended radio cocoons that are likely a result of some earlier activity period.
\item[-] High-resolution and high-sensitivity low-frequency observations with LOFAR or other low-frequency telescopes should help to find more diffuse cocoons around
radio galaxies and consequently to trace former periods of activity of these sources.
\end{itemize}

We note here that for other interferometric radio observatories at
higher frequencies (like the Square Kilometer Array, SKA), it may become essential to provide large single-dish antennas to supplement
the centres of their extended (u,v) planes.

\Acknow{The authors are grateful to Matteo Murgia for access to the SYNAGE software. M.J. and M.W. are supported by the Polish NSC Grant DEC-2013/09/B/ST9/00599. We 
also thank the anonymous referee for valuable comments that helped to improve this paper.}

\end{document}